\documentclass[twocolumn]{aastex631}
\renewcommand{\arraystretch}{1.2}
\usepackage{multirow, booktabs, tabularx}
\usepackage[para,online,flushleft]{threeparttable}
\usepackage{float}
\usepackage{graphicx} % Required for rotatebox
\graphicspath{ {./images/} }

\begin{document}

%\title{Multi-wavelength observations of a remarkable hyperluminous X-ray source-An IMBH outburst, or a changing-look AGN?}

\title{Multi-wavelength study of a hyperluminous X-ray source near NGC\,6099: a strong IMBH candidate}

\author[0009-0007-2115-6257]{Yi-Chi Chang}
\affiliation{Institute of Astronomy, National Tsing Hua University, Hsinchu 30013, Taiwan}

\author[0000-0002-4622-796X]{Roberto Soria}
\affiliation{INAF-Osservatorio Astrofisico di Torino, Strada Osservatorio 20, I-10025 Pino Torinese, Italy}
\affiliation{Sydney Institute for Astronomy, School of Physics A28, The University of Sydney, Sydney, NSW 2006, Australia}
\affiliation{College of Astronomy and Space Sciences, University of the Chinese Academy of Sciences, Beijing 100049, China}

\author[0000-0002-5105-344X]{Albert K.~H.~Kong}
\affiliation{Institute of Astronomy, National Tsing Hua University, Hsinchu 30013, Taiwan}

\author[0000-0002-6496-9414]{Alister W. Graham}
\affiliation{Centre for Astrophysics and Supercomputing, Swinburne University of Technology, Hawthorn, VIC 3122, Australia}

\author[0000-0003-3255-7340]{Kirill A. Grishin}
\affiliation{Universite Paris Cit\'e, CNRS, Astroparticule et Cosmologie, F-75013 Paris, France}
\affiliation{Sternberg Astronomical Institute, Moscow State University, 13 Universitetsky pr., Moscow, Russia}
%\affiliation{Universite Paris Cit\'e, CNRS, Astroparticule et Cosmologie, F-75013 Paris, France}

\author[0000-0002-7924-3253]{Igor V. Chilingarian}
\affiliation{Center for Astrophysics $\vert$ Harvard \& Smithsonian, Cambridge, MA 02138, USA}
\affiliation{Sternberg Astronomical Institute, Moscow State University, 13 Universitetsky pr., Moscow, Russia}

\email{judy.chang@gapp.nthu.edu.tw, akong@gapp.nthu.edu.tw}



\begin{abstract}
We report on the intriguing properties of a variable X-ray source projected at the outskirts of the elliptical galaxy NGC\,6099 ($d \approx 139$ Mpc).
If truly located near NGC\,6099, this is a hyperluminous X-ray source that reached an X-ray luminosity $L_{\rm X} \approx $ a few times $10^{42}$ erg s$^{-1}$ in 2012 February ({\it XMM-Newton} data), about 50--100 times brighter than in 2009 May ({\it Chandra}) and 2023 August ({\it XMM-Newton}). 
The X-ray spectrum was soft at all three epochs, with a thermal component at $kT \approx 0.2$ keV and a power-law photon index $>3$. 
Such properties make it a strong candidate for an intermediate mass black hole (IMBH).
We also discovered a point-like, blue optical counterpart ($m_{g,{\rm Vega}}\approx24.7$~mag, $M_{g,{\rm Vega}}\approx-11.2$~mag),
from images taken by the Canada-France-Hawaii Telescope, and later confirmed with {\it Hubble Space Telescope} observations. 
The optical continuum can be modeled as stellar emission from a compact star cluster or an X-ray-irradiated accretion disk, consistent with the IMBH scenario. 
We discuss alternative explanations for the nature of this system. 
A possible scenario is tidal stripping of an orbiting star, with repeated X-ray outbursts every few years. 
An alternative possibility is that the thermal X-ray emission seen in 2009 was from shocked gas in the self-intersecting tidal stream during the rising phase of a tidal disruption event, while the 2012 and 2023 emissions were from the fully-formed accretion disk.
\end{abstract}



\section{Introduction} \label{sec:intro}
The existence of intermediate mass black holes (IMBHs: \citealt{koliopanos17,Greene2020,volonteri21}) in the mass range $\sim$10$^3$--$10^4 M_\odot$ is favored by theoretical arguments, such as their role as seeds for the rapid growth of supermassive black holes (SMBHs) \citep{volonteri08,ricarte18,inayoshi20,larson23}, with masses $>$10$^9 M_\odot$ already at $z\gtrsim6$ \citep{mortlock11,banados18,yang20,wang21,fan23}.
Some of the unanswered questions are how many IMBHs have survived in the present-day universe, where they are located, and how we can observe them \citep{Mezcua_2017,barrows24}. 

Three possible local-universe environments where IMBHs may exist are: 
(a) in the core of massive globular clusters, where they may have been formed from the rapid core collapse and runaway merger of O stars \citep{miller02,Portegies_Zwart_2002,gurkan04,Freitag_2006,Di_Carlo_2021};
(b) in the nuclei of dwarf galaxies and late-type disk galaxies, if the scaling relations between spheroidal stellar mass and nuclear black hole (BH) mass can be extrapolated to such a mass range \citep{graham16,chilingarian18,davis21,Graham_2021,Graham_2023b}; 
(c) floating in the halo of massive galaxies (''wandering IMBHs'': \citealt{bellovary10,greene21,seepaul22,dimatteo23}), perhaps still inside a tightly bound stellar cluster, as a result of gravitational recoil and/or tidal stripping of accreted and disrupted satellite dwarfs. 
\par
In this work, we report on the X-ray spectral and timing properties of an IMBH candidate in the local universe: 2CXO\,J161534.2$+$192707 \citep{Evans_2010,Evans_2024} = 4XMM\,J161534.3$+$192707 \citep{Webb_2020,Tranin_2022}. 
This source is located in projection in the halo of the elliptical galaxy NGC\,6099 ($d \approx 139$ Mpc). 
If it is at the same distance as the galaxy, its X-ray luminosity was $L_{\rm X} \approx 4 \times 10^{42}$ erg s$^{-1}$ in 2012. We also discovered and investigated its point-like optical counterpart, seen at $m_{g,{\rm Vega}} \approx 24.7$ mag in 2022--2023. 
In Section 2, we describe the general reasons why we consider this source a strong off-nuclear IMBH candidate in the nearby universe, and we discuss its spatial association with the nearby galaxy.
In Section 3, we summarize the X-ray and optical observations used in this study, and our data analysis techniques.
In Section 4, we discuss the positional association of the X-ray source with a point-like optical counterpart.
In Section 5, we report on the brightness and colours of the optical counterpart.
In Section 6, we model the X-ray spectral properties of the source in the 2009 {\it Chandra} observation (lower state), the 2012 {\it XMM-Newton} observation (higher state), and the 2023 {\it XMM-Newton} observation (lower state again). 
In Section 7, we combine the optical and X-ray data and model the broad-band spectral energy distribution (SED); we illustrate the case of an irradiated accretion disk and of a star cluster.
In Section 8, we summarize the multiband results and discuss possible interpretations for the source and its X-ray flux evolution.

\section{X-ray search of IMBH candidates}
\subsection{General selection criteria}
\label{section:selection_criteria}
Our focus is on IMBHs that are (or briefly become) X-ray active because of gas accretion, and that are not spatially coincident with the nucleus of a major galaxy. 
Unfortunately, it is generally hard to distinguish between an off-nuclear X-ray bright IMBH, a stellar mass X-ray binary and a background AGN that only happens to be seen in projection behind a nearby galaxy \citep[{\it{e.g.}},][]{Zolotukhin2016,earnshaw19,barrows24}. 
\par
Our first selection criterion for a point-like, non-nuclear X-ray source to be considered an IMBH candidate is that it is probably too luminous to be a stellar-mass compact object at the distance of its apparent host galaxy. 
Stellar-mass BHs and neutron stars can reach apparent luminosities of several times $10^{40}$ erg s$^{-1}$, at highly super-Eddington accretion rates, and with the help of moderate polar-funnel beaming (ultraluminous X-ray sources, ULXs: \citealt{2011NewAR..55..166F,Kaaret_2017,king23}).
Thus, a somewhat conventional threshold to screen out stellar-mass accretors is an apparent isotropic 0.3--10 keV luminosity of $10^{41}$ erg s$^{-1}$: sources above this limit are usually referred to as hyperluminous X-ray sources (HLXs) \citep{gao03,king05,sutton12,Barrows_2019,mackenzie23,tranin24}. 
\par
An apparent spatial association with a nearby galaxy does not guarantee that an X-ray source is really at the same distance as the galaxy: in the absence of redshift information on its optical counterparts, it may also be a background AGN randomly projected behind the galaxy \citep{masetti03, heida13,sutton15, earnshaw19}. 
To reduce the effect of this contamination, we searched for HLXs in the thermal dominant (''high/soft'') state \citep{maccarone03,fender04,Remillard_2006}, that is at a dimensionless accretion rate $0.02 \lesssim \dot{m} \lesssim 1$, where $\dot{m} \equiv \dot{M}/\dot{M}_{\rm Edd}$ and $\dot{M}_{\rm Edd}$ is the Eddington accretion rate, with $\dot{M}_{\rm Edd} \equiv L_{\rm Edd}/ 0.1c^2$, and $L_{\rm Edd}$ is the Eddington luminosity.
We assume that the high/soft state of an accreting IMBH behaves like that of a stellar-mass BH, dominated by a standard Shakura-Sunyaev multicolor disk spectrum \citep{Shakura&Sunyaev_1973,merloni00} with peak color temperature $kT_{\rm in} \approx 230 \left( \dot{m}/M_4 \right)^{1/4}$ eV \citep{Shakura&Sunyaev_1973, Kubota_1998, Done_2012}, where $M_4$ is the BH mass in units of $10^4 M_{\odot}$. 
Therefore, point-like X-ray sources with apparent luminosities $\sim10^{41}$--$10^{43}$ erg s$^{-1}$, a soft thermal spectrum, and peak disk-blackbody temperatures $kT_{\rm in} \approx 0.2$--0.3 keV are strong IMBH candidates. 
The vast majority of background AGN (powered by SMBHs) have a more dominant power-law component above 1 keV, with characteristic photon indices $\approx$1.0--2.5 \citep{piconcelli05,page06,corral11,marchesi16,kynoch23,elias-chavez24}, plus a fainter thermal component (''soft excess'') with characteristic temperatures $kT \approx 50$--150 eV \citep{Crummy_2006,Done_2012}, cooler than expected from an IMBH.
\par
A further IMBH selection criterion is an X-ray flux variability of at least an order of magnitude between separate epochs observed by either {\it Chandra} or {\it XMM-Newton}. 
This is because an off-nuclear IMBH is unlikely to accrete from a steady galaxy-scale gas inflow, or to receive a steady Roche-lobe mass transfer from a much less massive donor star. 
Either in the case of accretion from an individual donor star \citep{kalogera04}, or in the case of feeding from a tidal disruption event (TDE) \citep{stone16,Lin_2018,saxton20}, we expect transient X-ray outbursts. 
\par
Finally, an X-ray over optical flux ratio $F_{\rm X}/F_{\rm opt} \gtrsim 10$ provides additional support to the IMBH classification. 
Most AGNs have $0.1 \lesssim F_{\rm X}/F_{\rm opt} \lesssim 10$ \citep{lusso10,civano12,heida13}, because their emission from the accretion disk peaks in the UV band. 
Foreground Galactic stars have a soft X-ray spectrum, but $F_{\rm X}/F_{\rm opt} \lesssim 10^{-3}$. 
In practice, it is often difficult to have simultaneous X-ray and optical coverage of a source, especially in the case of X-ray transients.
\par
The criteria outlined above are not a silver bullet for IMBH detection. 
For example, a small fraction of AGN (''supersoft AGN'') have a thermal spectrum with temperatures as high as $\sim$0.2 keV and almost no power-law component or a power-law with photon idex $\Gamma > 3$  \citep{Terashima_2012,sacchi23}, similar to the predicted spectrum of an IMBH in the high/soft state. 
Nuclear SMBHs that become active as a result of a TDE are transient and often have purely thermal X-ray emission \citep{saxton20,Gezari_2021}. 
Other AGN, known as changing-look or changing-state AGN \citep{Komossa_2023}, show X-ray luminosity variations by up to two orders of magnitude over a few years, sometimes also accompanied by a soft, thermal X-ray spectrum \citep{Masterson_2022}. 
Moreover, at least one confirmed stellar-mass accretor, the ULX pulsar in NGC\,5907, has approached an X-ray luminosity $\approx$10$^{41}$ erg s$^{-1}$ \citep{israel17,furst17}.
Nonetheless, searching for soft X-ray transients in the outskirts of early-type galaxies, at $L_{\rm X} \sim 10^{41}$--$10^{43}$ erg s$^{-1}$, does provide a useful first screening of IMBH candidates for detailed follow-up investigations.
\par
We applied the selection criteria described above to the {\it XMM-Newton} sources in the catalog of \cite{Tranin_2022}. 
Specifically, we selected sources with: i) a luminosity $L_{\rm X} > 10^{41}$ erg $s^{-1}$ if located in the apparent host galaxy; ii) a luminosity distance of the presumed host galaxy $<$300 Mpc; iii) a photon index of the best power-law spectral fit $\Gamma > 3.0$; iv) EPIC-pn hardness ratios\footnote{The hardness ratio HR3 is defined as $(F3-F2)/(F3+F2)$, where $F3$ is the observed EPIC-pn flux in the 2.0--4.5 keV band and $F2$ is the flux in the 1.0--2.0 keV band. The hardness ratio HR4 is defined as $(F4-F3)/(F4+F3)$, where $F4$ is the 4.5--12.0 keV flux and F3 is the 2.0--4.5 keV flux. The softer a source is, the closer both those ratios are to $-1$.} HR3 $< -0.5$ and HR4 $< -0.5$; v) at least two observations between {\it XMM-Newton}, {\it Chandra} and {\it Swift}, with flux variations larger than 10 between different epochs.

\begin{figure*}
    \includegraphics[width=2.12\columnwidth]{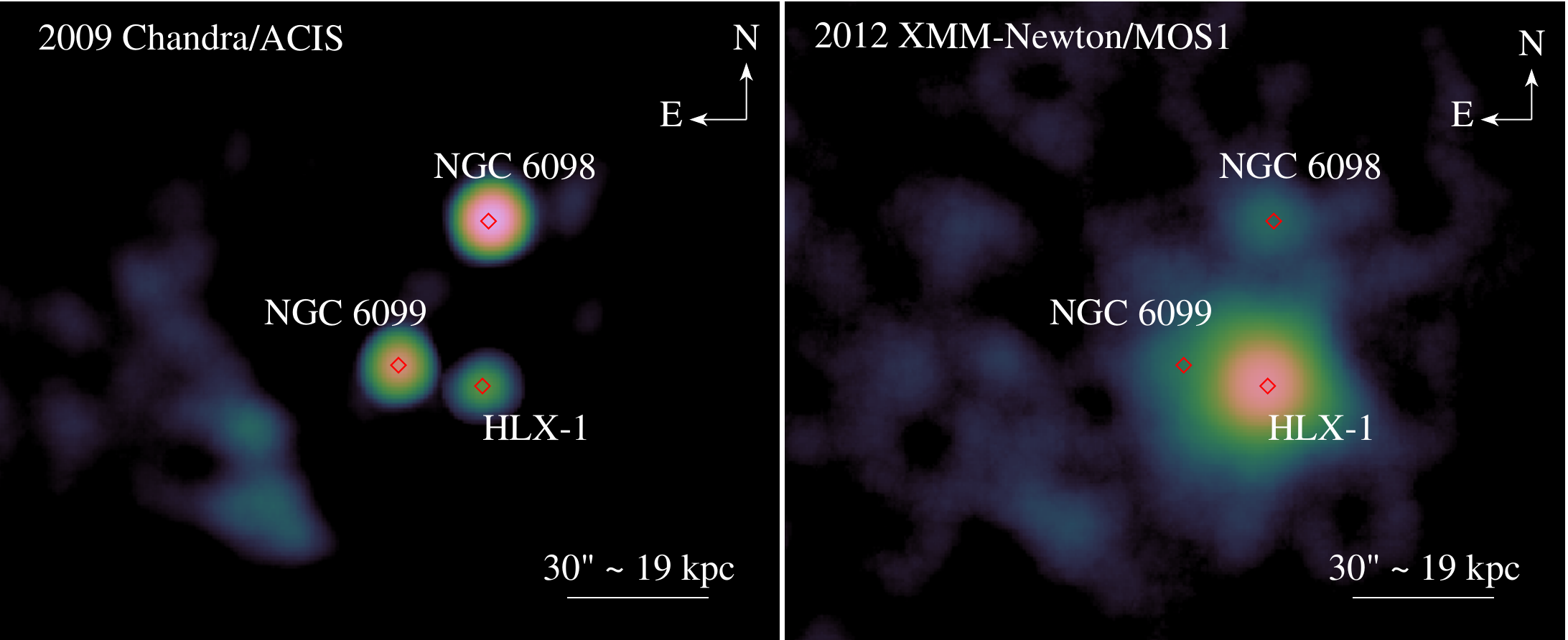}\\
    \includegraphics[width=1.05\columnwidth]{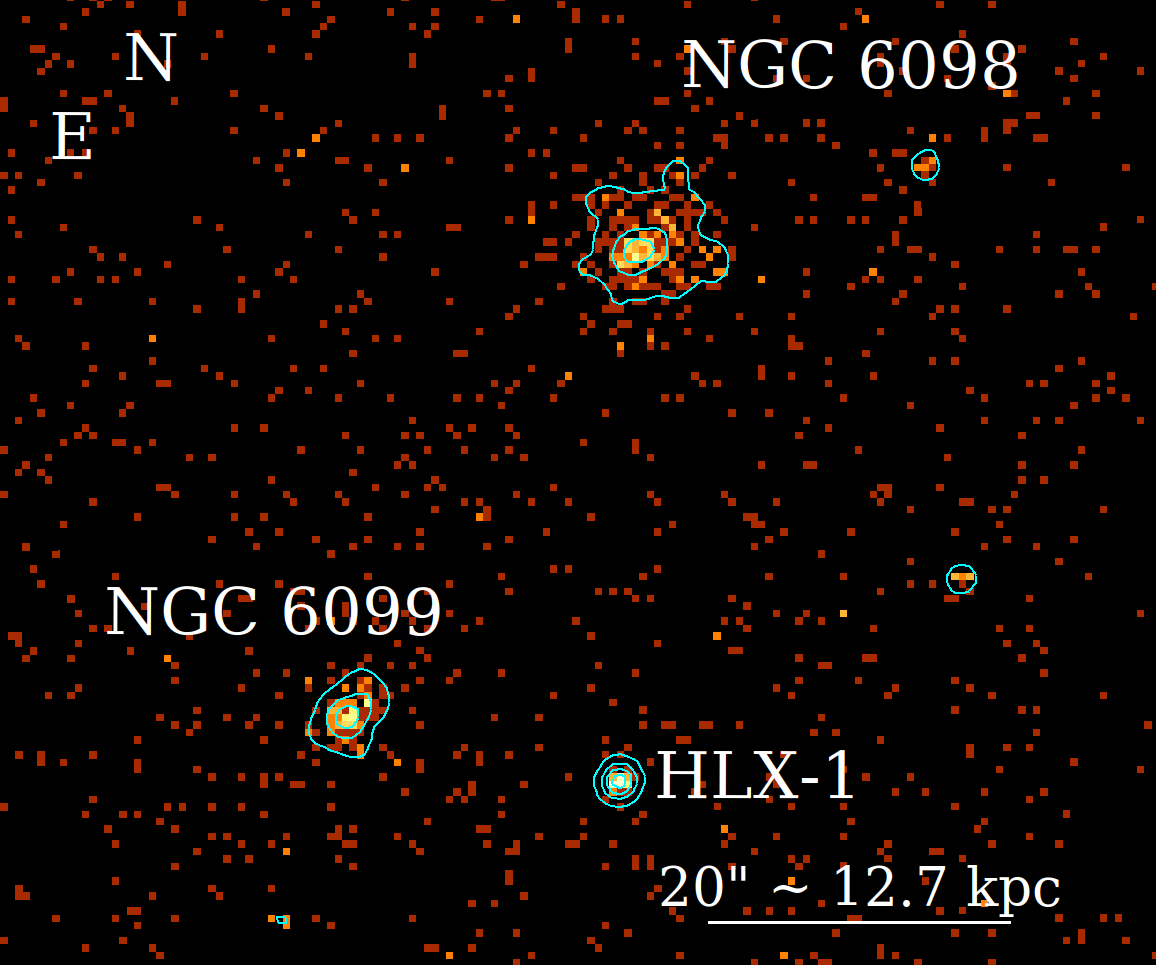}
    \includegraphics[width=1.05\columnwidth]{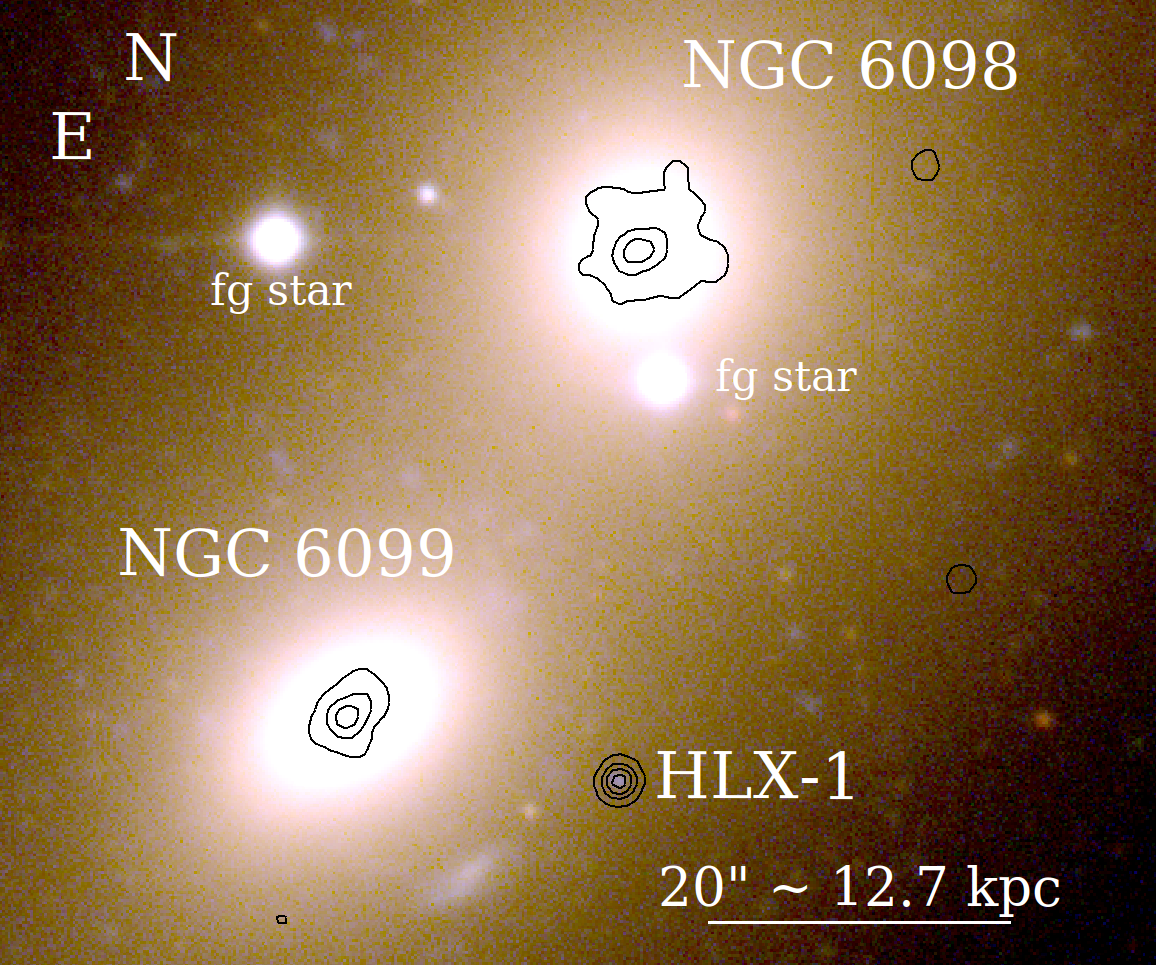}
    \caption{Top left: false-color, adaptively smoothed image from the 2009 {\it Chandra}/ACIS observation (0.3--8 keV band). Here, HLX-1 has an X-ray luminosity of a few $\times 10^{40}$ erg s$^{-1}$, slightly fainter than its two neighboring galaxies. A ridge of diffuse emission is also visible, $\approx$20 kpc south-east of NGC\,6099. Top right: false-color, adaptively smoothed {\it XMM-Newton}/EPIC-MOS1 image (0.3--10 keV band) from 2012, displayed on the same spatial scale as the top left panel.
    At this epoch, HLX-1 vastly outshines the X-ray emission from NGC\,6099 and NGC\,6098.
    Bottom left: zoomed-in, unsmoothed 2009 {\it Chandra}/ACIS image (0.3--8 keV band), with smoothed X-ray flux contours overplotted in cyan (linear scale). Bottom right: true-color CFHT/Megacam image on the same scale as the bottom left panel, with {\it Chandra}/ACIS flux contours overplotted in black. The optical colors are blue = $u$ band, green = $g$ band, and red = $r$ band. The blue-ish counterpart of HLX-1 stands out from the old-population halo of NGC\,6099 and NGC\,6098.}
    \label{fig:fig1}
\end{figure*}

\subsection{Our first cut of interesting soft sources}
\label{section:tranin_catalog}
Out of about half a million sources in the \cite{Tranin_2022} catalog, only seven sources satisfy the five selection criteria detailed above. 
We inspected and investigated each of these sources (using archival data and literature results) to check whether they did contain plausible IMBH candidates. 
One of the seven sources is an overdensity of hot gas in the nuclear region of the elliptical galaxy NGC\,741 \citep{jetha08,schellenberger17}, and was therefore immediately dismissed. 
Two sources are well-known supersoft, strongly variable AGN: the narrow line Seyfert galaxies 1H 0707$-$495 = LEDA 88588 \citep{boller02,dauser12,hancock23,xu24,dobrotka24} and IRAS 13224$-$3809 = LEDA 88835 \citep{jiang18,pinto18,alston19,caballero20,jiang22,hancock23,xu24,dobrotka24}. 
Both AGN are of course extremely interesting objects for studies of ultrafast outflows, Compton reflection spectra, disk flares; but they are not relevant to our search of off-nuclear IMBH candidates. 
A further source is associated to a close pair of merging galaxies at the outskirts of the galaxy cluster Abell 1795: the LINER 2MASX J13482545$+$2624383 and the elliptical galaxy SDSS J134825.62$+$262435.7 \citep{nisbet16,abdullah20}. 
The large distance ($\approx$280 Mpc) of Abell 1795 makes it impossible to determine whether the soft {\it XMM-Newton} source corresponds to the nucleus of SDSS J134825.62$+$262435.7 or is slightly offset; an analysis of this system is left for further work.
\par
This leaves three sources in our selected sample. 
One of them is ESO 243-49 HLX-1 \citep{Farrell2009,soria17}, projected in the halo of an S0 galaxy ($d \approx 95$ Mpc). 
It is perhaps the best-known off-nuclear IMBH candidate in the literature; it has shown repeated outbursts with state transitions, and has a thermal disk-blackbody spectrum at the peak of each outburst, with a peak luminosity $L_{\rm X} \approx 10^{42}$ erg s$^{-1}$. 
The second one is the other well-known IMBH candidate 3XMM J215022.4$-$055108 \citep{Lin_2018}. 
It is a soft, thermal, transient source (probably triggered by a TDE) in a globular cluster at the outskirts of a lenticular galaxy ($d \approx 250$ Mpc). 
Its maximum observed luminosity is $L_{\rm X} \approx 10^{43}$ erg s$^{-1}$, although it may have reached an even higher luminosity at the (unobserved) outburst peak. 
This leaves us with the last one of those seven soft, variable X-ray sources: 2CXO\,J161534.2$+$192707 = 4XMM\,J161534.3$+$192707, near the elliptical galaxy NGC\,6099 (Figure~\ref{fig:fig1}). 
Henceforth, we refer to this source as NGC\,6099 HLX-1 or simply HLX-1, for simplicity. 
It has received no special attention in the literature so far\footnote{The same source 2CXO\,J161534.2$+$192707 is also listed in the ULX catalogs of \cite{kovlakas20} and \cite{Bernadich_2022}; however, the host galaxy identification and distance are wrong (it was mistakenly associated with a foreground star) and therefore it was not recognized as an HLX.}. 
We will show in this work that it is a strong IMBH candidate and that it has a point-like optical counterpart.

%\subsection{Our best IMBH candidate in the halo of NGC\,6099}
\subsection{\texorpdfstring{Our best IMBH candidate in the halo of NGC\,6099}{Our best IMBH candidate in the halo of NGC 6099}}
NGC\,6099 HLX-1 is projected inside the elliptical galaxy NGC\,6099, $\approx$18\farcs3 west of its core. 
The $r$-band Petrosian radius \citep{Petrosian1976} of NGC\,6099 is 14\farcs6 \citep{Ahumada_2020}\footnote{By comparison, the $B$-band D25 isophotal diameter listed in the NASA/IPAC Extragalactic Database (NED, \url{https://ned.ipac.caltech.edu}) is 79\farcs1, because it refers to the 25-mag isophote around the galaxy pair NGC\,6098/99.}. 
Two Petrosian radii can be taken as a conventional size to define the association of an X-ray source with a host galaxy \citep{graham05,Zolotukhin2016}.
\par
NGC\,6099 has a spectroscopic redshift $z = 0.03043 \pm 0.00001$ \citep{Ahumada_2020}, corresponding to a luminosity distance $d = (139 \pm 10)$ Mpc for the standard cosmological parameters adopted by NED ($H_0 = 67.8$ km s$^{-1}$ Mpc$^{-1}$, $\Omega_m = 0.308$, $\Omega_\Lambda = 0.692$).
It has a gravitationally bound, close companion with a similar size, morphology and optical brightness, NGC\,6098\footnote{Together, they form the galaxy pair KPG 493.} (Figure~\ref{fig:fig1}), with a redshift $z = 0.03155 \pm 0.00001$ corresponding to a luminosity distance $d = (144 \pm 10)$ Mpc.
It is likely that the discrepant redshift measured for the two elliptical galaxies corresponds to their relative proper motion ($\approx$330 km s$^{-1}$)\footnote{The relative motion of the two galaxies may be responsible for the ridge of soft, thermal X-ray emission from shocked gas, $\approx$30 kpc to the south-east of the nucleus of NGC\,6099, visible in both {\it XMM-Newton} and {\it Chandra} observations. An analysis of the properties of the shocked gas is beyond the scope of this paper.} around their center of mass; in this case, the true luminosity distance of both galaxies may be intermediate between the two values given above. 
For this work, in the absence of any information on the proper motion of the two galaxies, we adopt the luminosity distance of 139 Mpc for NGC\,6099 HLX-1, tied to the luminosity distance of its apparent host galaxy.
At this distance, HLX-1 reached $L_{\rm X} \sim $ a few $10^{42}$ erg s$^{-1}$ in one observation, and varied in flux by a factor of at least 50 between observations (Figure~\ref{fig:fig1}).

\section{Observations and data analysis}
\subsection{Chandra X-ray Observatory (2009)}
The field of NGC\,6098/6099 was observed with the Advanced CCD Imaging Spectrometer (ACIS) onboard NASA's {\it Chandra X-ray Observatory}, on 2009 May 23, for a good time interval (GTI) of 44.6 ks (Table~\ref{tab:obs_info}).
We retrieved the observation files from the {\it Chandra} Data Archive, and reprocessed them with the {\it Chandra} Interactive Analysis of Observations ({\sc ciao}) version 4.13 \citep{Fruscione_2006}, with the Calibration Database (CALDB) version 4.9.8. 
We created new level-2 event files with the {\sc ciao} task {\fontfamily{qcr}\selectfont chandra\_repro}. 
We checked with {\fontfamily{qcr}\selectfont dmextract} that there were no significant background flares during the observation. 
We used {\fontfamily{qcr}\selectfont dmcopy} for energy filtering and to create images in different bands, and the imaging tool {\sc ds9} to display the images and define source and background regions.
For the point-like source HLX-1, the source extraction region was a circle with a 2$\arcsec$ radius; the background region was an annulus with an inner radius of 3.5$\arcsec$ and an outer radius of 7$\arcsec$.
For the diffuse emission in the nuclei of the two elliptical galaxies, we used circular source regions with a radius of 4$\arcsec$ for NGC\,6099 and 5$\arcsec$ for NGC\,6098; the associated background regions were chosen to be at least 4 times the source extraction areas, and not including X-ray photons from HLX-1 or from other galaxies. 
We then used {\fontfamily{qcr}\selectfont specextract} to extract a spectrum and associated response and ancillary response files for HLX-1 (option {\fontfamily{qcr}\selectfont correctpsf} = yes), and for the extended emission in the nuclear regions (option {\fontfamily{qcr}\selectfont correctpsf} = no). 
The arcsec resolution of {\it Chandra} guarantees that there is no significant contamination of the ACIS spectrum of HLX-1 with diffuse thermal emission from NGC\,6098 and NGC\,6099, and vice versa (Figure~\ref{fig:fig1}). 
\par
For our spectral analysis, we used the {\sc ftools} \citep{Blackburn_1995,NASA_2014} software suite, from NASA’s High Energy Astrophysics Science Archive Research Center (HEASARC).
We rebinned the spectra to $\ge$1 count per bin with {\fontfamily{qcr}\selectfont grppha}, and modelled them in the 0.3--8.0 keV band with the spectral analysis package {\sc xspec} version 12.13.0c \citep{Arnaud_1996}. 
Because of the low number of counts (only $\approx$50 net counts for HLX-1), we used the {\fontfamily{qcr}\selectfont cstat} statistics \citep{Cash_1979} for the fit statistics. 
Error ranges (90\% confidence limit) for the fit parameters were calculated with the {\fontfamily{qcr}\selectfont {steppar}, {error}} task in {\sc xspec}. 
Confidence limits for the luminosity distances were calculated with the {\sc Python} package {\fontfamily{qcr}\selectfont {uncertainties}} version 3.1.7, {\url{http://pythonhosted.org/uncertainties/}}.

\begin{figure*}
	% Allowable file formats are eps or ps if compiling %using latex or pdf, png, jpg if compiling using pdflatex
    \centering
    \includegraphics[width=1.745\columnwidth]{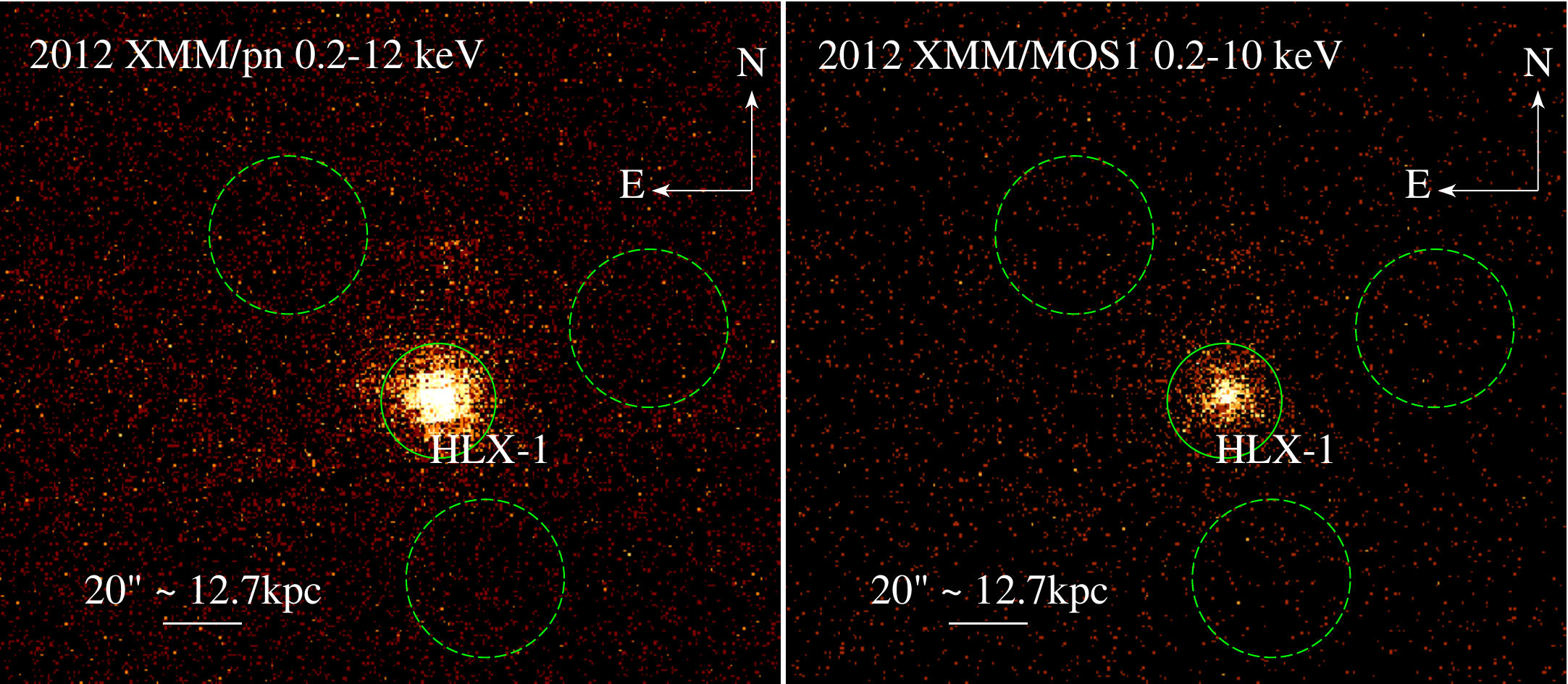}\\
    \caption{{\it XMM-Newton}/EPIC pn (left panel) and MOS1 (right panel) images from the 2012 observation. The solid green circles (radius of 14\farcs5) define the source extraction region for our timing and spectral analysis of HLX-1; the dashed circles (radii of 20\arcsec) map the background regions. 
    The contribution from the diffuse emission around the nuclei of NGC\,6098 and NGC\,6099 is negligible, compared with the emission from HLX-1.
}
    \label{fig:XMM1-image}
\end{figure*}

\begin{figure*}
	% Allowable file formats are eps or ps if compiling %using latex or pdf, png, jpg if compiling using pdflatex
    \centering
    \includegraphics[width=1.745\columnwidth]{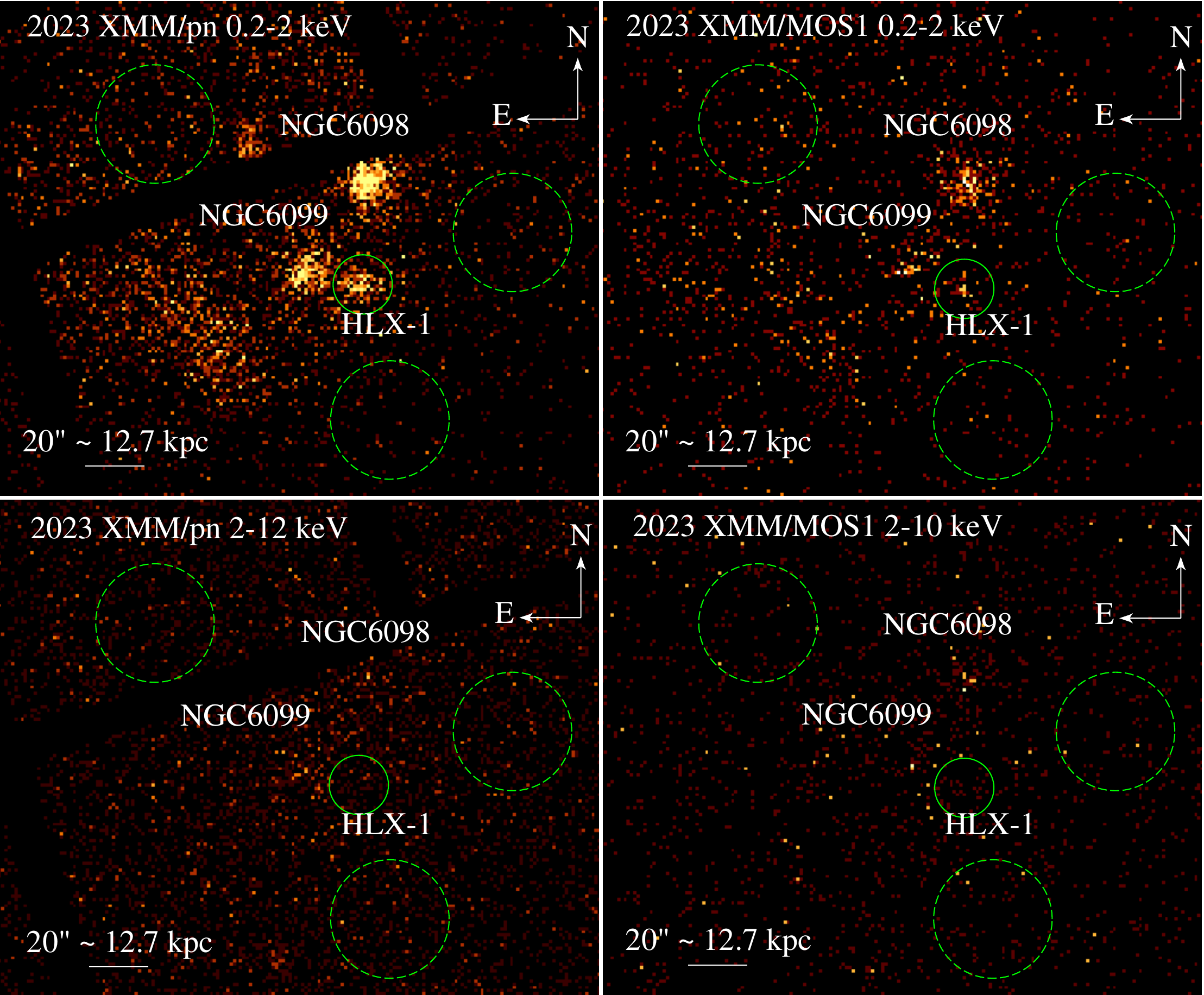}\\
    \caption{{\it XMM-Newton}/EPIC pn and MOS1 images from the 2023 observation, in different energy bands; the field of view is the same as in Figure~\ref{fig:XMM1-image}. Top row: soft band (0.2--2 keV); bottom row: hard band (2--12 keV for pn, 2--10 keV for MOS1). 
    In each panel, the solid green circle (radius of 10\arcsec) is the source extraction region for HLX-1, and the dashed green circles (radius of 20\arcsec) map the background regions. These images show that the 2023 flux of HLX-1 has declined below the flux of the diffuse emission from NGC\,6099; it also highlights the softness of its spectrum.
    }
    \label{fig:XMM2-image}
\end{figure*}

\begin{table}
\centering
    \caption{Log of the X-ray observations of the field of NGC\,6099 analysed in this paper.}
    \begin{tabular}{*{5}{c}}% It allows you to set the width of the table and provides the 7 column type, which fills out the rest of the space and alignment for each. 
        \hline
        Observ. & Detector & ObsID & Start Date  & GTI \\
            &         &            &         &   (ks) \\
        \hline
        Chandra & ACIS &  10230    & 2009 May 23  &  44.6  \\
        XMM1 & MOS & 0670350301  & 2012 Feb 25 &  14.5 \\
             & pn & 0670350301  & 2012 Feb 25 &   10.6 \\
        Swift1 & XRT & 00014346001 & 2021 Jun 18 &    1.7  \\
        Swift2 & XRT & 00014346002 & 2021 Jun 20 &   0.9   \\
        Swift3 & XRT & 00097139001 & 2023 Apr 9 &    2.8   \\
        Swift4 & XRT & 00097139002 & 2023 Jun 28 &  1.8   \\
        Swift5 & XRT & 00097139003 & 2023 Jul 12 &    1.0   \\
        XMM2  & MOS & 0923030101  & 2023 Aug 4 & 85.2  \\
               & pn & 0923030101  & 2023 Aug 4 & 69.7  \\
        Swift6 & XRT & 00097139004 & 2023 Oct 9 &  2.5   \\
        Swift7 & XRT & 00097139005 & 2024 Jan 9 &  2.8 \\
        \hline
    \end{tabular}
    \label{tab:obs_info}
\end{table}

\subsection{XMM-Newton (2012, 2023)}
The European Space Agency's {\it XMM-Newton} space observatory observed the field of NGC\,6098/6099 on two occasions: on 2012 February 25 (ObsID 0670350301, Rev.\ 2237, PI: John Mulchaey), with an exposure time of $\approx$16.6 ks, and on 2023 August 4 (ObsID 0923030101, Rev.\ 4332, PI: Albert Kong), for $\approx$88 ks. 
Henceforth, we will usually refer to those two datasets as XMM1 and XMM2, for shorthand notation. 
On both occasions, the European Photon Imaging Camera (EPIC) was the prime instrument, with two MOS and one pn detector. 

We downloaded the data files from the {\it XMM-Newton} Science Archive; XMM1 was already in the public domain, while XMM2 was the result of a follow-up observation proposed by our team specifically for HLX-1; it is also now publicly available.
We processed both datasets with the {\it XMM-Newton} Science Analysis System ({\sc sas}) v20.0.0 software, with the calibration files updated to 2021 November 13. 
We used the {\fontfamily{qcr}\selectfont epproc} and {\fontfamily{qcr}\selectfont emproc} tasks to reprocess observation data files (ODFs) and obtain calibrated and concatenated event lists. 
To filter out intervals of flaring background, we set a rejection threshold of 0.35 ct s$^{-1}$ for each MOS and 0.4 ct s$^{-1}$ for the pn, over the full field of view at channel energies above 10 keV; the filtering was done with the {\sc sas} task {\fontfamily{qcr}\selectfont tabgtigen}. 
In XMM1, we retained a GTI of 14.5 ks for EPIC-MOS and 10.6 ks for EPIC-pn; in XMM2, the GTI was 85.2 ks for EPIC-MOS and 69.7 ks for EPIC-pn.
\par
We used the {\sc sas} task {\fontfamily{qcr}\selectfont xmmselect} to define source and background extraction regions, and to build spectral files and associated response and ancillary response files, for each detector.
Given the small angular separation between HLX-1 and the nucleus of NGC\,6099 ($\approx$18$\farcs$3, computed from the {\it Chandra} image), we need to assess and minimize the possible contamination of the HLX-1 spectrum from the diffuse hot gas emission of the nearby galaxy. 
This means taking a source extraction region for HLX-1 large enough to include most of the source counts, but not so large to be significantly contaminated by the diffuse NGC\,6099 emission. 
The latter can be assumed to be constant over human timescales. 
Its intensity was measured from the {\it Chandra} data (Section~\ref{section:Chandra-obs}) and converted to the expected {\it XMM-Newton}/EPIC count rates with the Portable, Interactive Multi-Mission Simulator ({\sc pimms}) software, version 4.12d. 
Its spatial extent in {\it Chandra}/ACIS was then convolved with the point spread function (PSF) of the {\it XMM-Newton}/EPIC detectors.
\par
We find that in XMM1, the total diffuse NGC\,6099 emission is insignificant compared with the emission from HLX-1 (Figure~\ref{fig:XMM1-image}). 
For all three EPIC instruments of the XMM1 dataset, we used a source extraction circle of $14\farcs5$ radius for HLX-1, and estimated that the contamination of galactic hot gas emission falling inside that circle is $<$1\%. 
For XMM2, the total diffuse NGC\,6099 emission is slightly higher than the HLX-1 emission (Figure~\ref{fig:XMM2-image}).
Thus, we chose to reduce the HLX-1 source extraction circle to a radius of $10\arcsec$, to keep the contamination from diffuse galactic emission to $<$1\% of the HLX-1 source emission.
On the contrary, any possible contamination from the diffuse emission in the other nearby elliptical galaxy, NGC\,6098, is completely negligible both in 2012 and in 2023. 
For both the XMM1 and XMM2 datasets, we defined the composite background region as three circles, each with a radius of 20$\arcsec$, within $\approx$1$\arcmin$ of the HLX-1 position, taking care to avoid the nuclear emission from NGC\,6098 and NGC\,6099, the diffuse emission ridge south-east of NGC\,6099, and any chip gaps. 
\par
With this choice of source and background regions, we extracted EPIC-pn lightcurves of HLX-1, with the standard tasks {\fontfamily{qcr}\selectfont xmmselect} and {\fontfamily{qcr}\selectfont epiclccorr}. 
We then extracted EPIC-MOS and EPIC-pn spectra and built the associated response and ancillary response files with {\fontfamily{qcr}\selectfont xmmselect}, {\fontfamily{qcr}\selectfont rmfgen} and {\fontfamily{qcr}\selectfont arfgen}. 
For all lightcurves and spectra, we used a spectral bin size of 5 eV.
\par
We rebinned the EPIC-pn and EPIC-MOS spectra to a minimum of 20 counts per bin with {\fontfamily{qcr}\selectfont grppha}, suitable for the $\chi^2$ fitting statistics. 
For each observation dataset, we modelled the three EPIC spectra, simultaneously with {\sc xspec}.
As a first step, we selected the energy range of 0.2--10 keV for the MOS detectors and 0.2--12 keV for the pn, for the 2012 and 2023 datasets. 
However, the 2023 images (Figure~\ref{fig:XMM2-image}) and spectra show no significant net counts from HLX-1 above 5 keV, and very few counts above 2 keV. 
Therefore, to limit the contamination from background photons, we ignored channels above 5 keV in the 2023 spectra.
We used the $\chi^2$ statistics for the computation of the best-fitting parameter values and 90\% confidence intervals of the upper and lower bounds.

\subsection{Swift X-Ray Observatory}
The X-Ray Telescope (XRT) on board the {\it Niels Gehrels Swift Observatory} took seven snapshot observations of NGC\,6099 between 2021 June 18 and 2024 January 9. 
The total effective ({\it{i.e.}}, accounting for vignetting) exposure time at the location of HLX-1 was 12.3 ks. 
We retrieved the data from NASA's HEASARC public archive, and processed them with the {\it Swift}/XRT data analysis tools available online\footnote{\url{https://www.swift.ac.uk/user_objects/}}.
X-ray emission from the nuclei of NGC\,6098 and NGC\,6099 is marginally detected in the stacked 0.3--10 keV image, but HLX-1 is not detected in any observation, nor in the stacked dataset. 
We estimated the total observed counts in a 10$\arcsec$ source extraction circle in the stacked image, and the expected background counts in regions at a similar distance from the nucleus of NGC\,6099. 
We then applied the statistics of \cite{Kraft_1991} to determine the confidence interval for a source with a low number of counts. 
After taking into account the XRT aperture correction from 10$\arcsec$ to infinity, we converted the 90\% upper limit on the net source counts into a 90\% upper limit on the observed flux, assuming the same spectral model fitted to the nearest {\it XMM-Newton} observations (XMM2). 
With this method, and correcting for line-of-sight absorption, we estimate a 90\% upper limit to the luminosity of HLX-1 in the stacked {\it Swift}/XRT dataset of $L_{\rm X} = (5 \pm 1) \times 10^{40}$ erg s$^{-1}$, consistent with the {\it XMM-Newton} results from 2023 (Section~\ref{section:XMM2_obs}).

\subsection{Canada-France-Hawaii Telescope}
We observed the field of NGC\,6099 with the MegaPrime/MegaCam detector on the 3.6-m Canada-France-Hawaii Telescope (CFHT) on 2022 August 2 (OBSID 2773230).
The images cover a $1^{\circ} \times 1^{\circ}$ field of view, with a resolution of 0\farcs187 per pixel.
We took four images in the {\it u} band (u.MP9302 filter), two in the {\it g} band (u.MP9402 filter) and four in the {\it r} band (u.MP9602 filter). 
The total exposure time was 2800 s for {\it u} and {\it r}, and 1200 s for {\it g}. 
The dimm mode seeing varied from 0$\farcs$51 to 0$\farcs$72.
Data were pre-processed for overscan modelling, bias subtraction, bad-pixel masking and flat-fielding, with the {\sc elixir} software \citep{Gwyn_2008}. 
We then used the {\sc iraf} task {\fontfamily{qcr}\selectfont imcombine} to create three combined images in the three bands, with cosmic rays filtered out.

\subsection{Hubble Space Telescope (HST)}
The field of NGC\,6099 was also observed by the Wide Field Camera 3 (WFC3) on board {\it HST}, on 2023 September 5 (PI: I.\ Chilingarian). Images were taken in three UVIS bands: F300X (exposure time of 696 s), F475W (696 s) and F814W (648 s). 
We downloaded the drizzled Single Visit Mosaics provided as Hubble Advanced Products (HAP-SVM), from the Barbara A.\ Mikulski Archive for Space Telescopes (MAST).
Detailed analysis and results of those observations will be reported elsewhere (K.\ Grishin et al., in prep.). 
In this paper, we use preliminary results taken from that forthcoming work, to constrain the physical size (Section~\ref{section:astrometry}) and brightness (Section~\ref{section:photometry}) of the optical counterpart and to improve our X-ray/optical spectral modelling (Section~\ref{section:SED}). 

\begin{figure}[t]
    \includegraphics[width=\columnwidth]{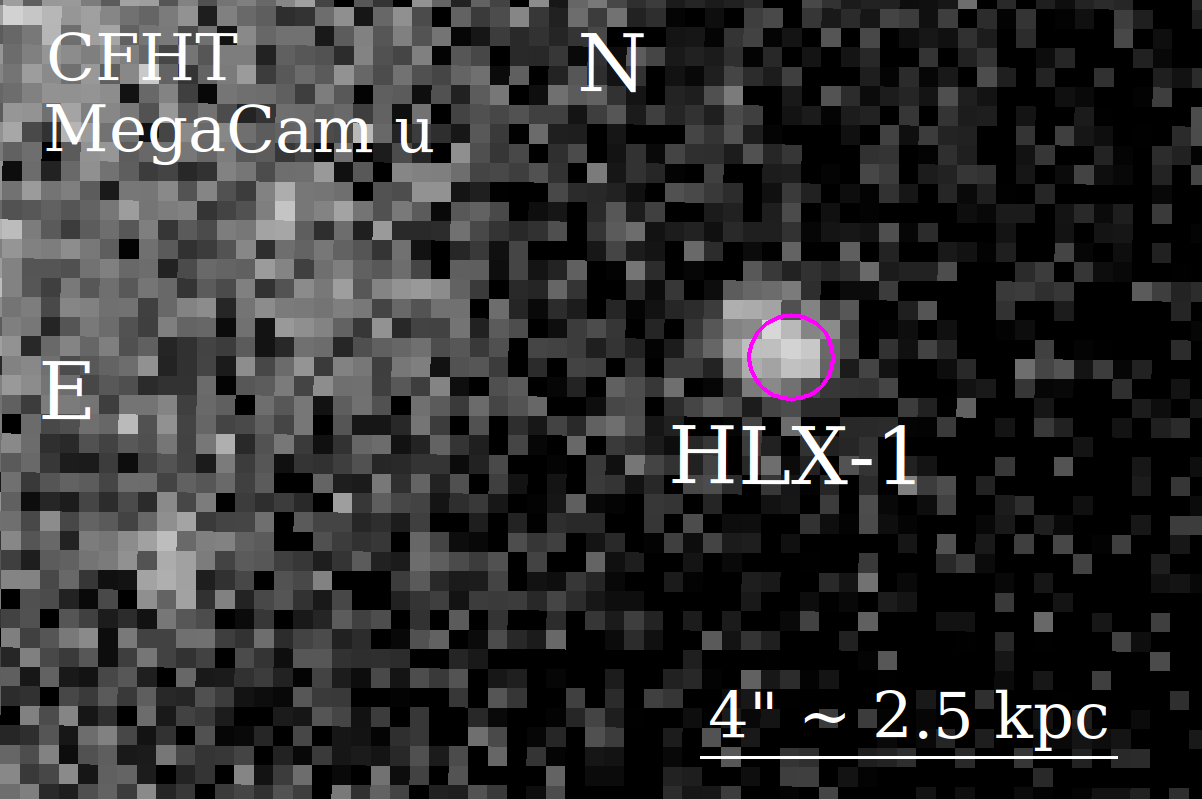}\\
    \includegraphics[width=\columnwidth]{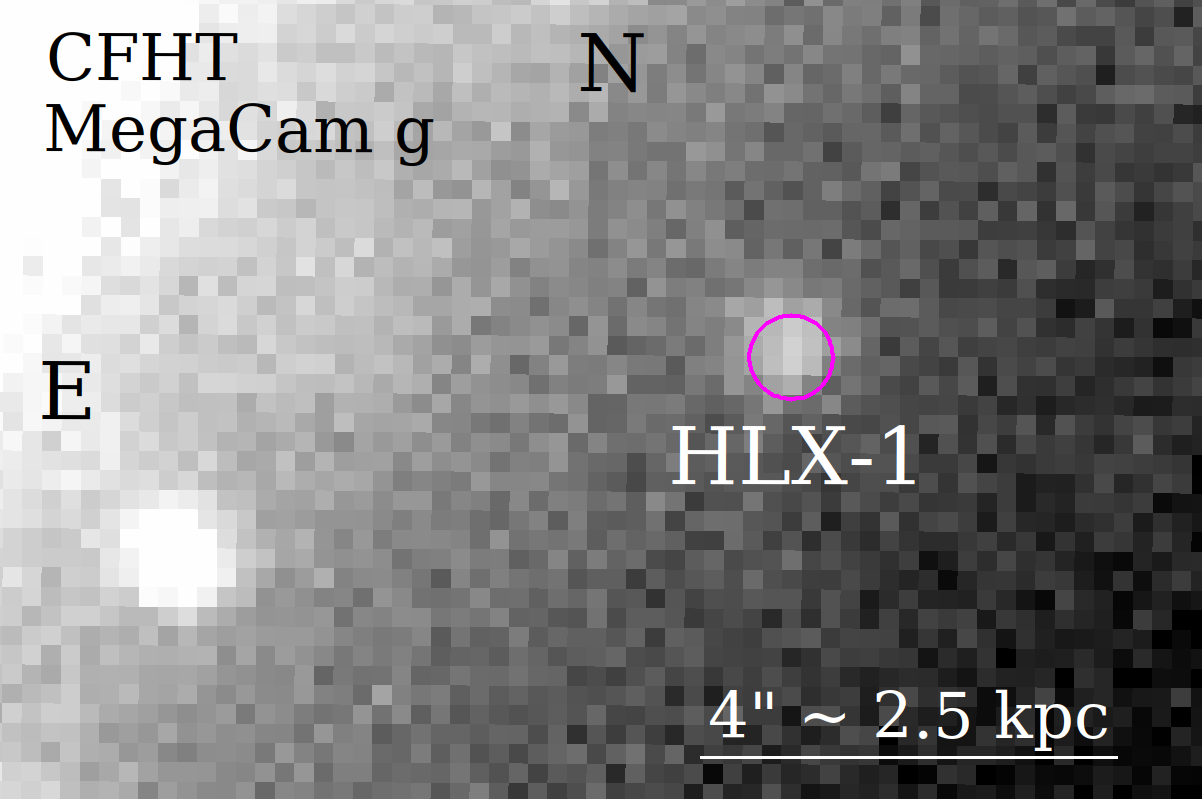}\\
    \includegraphics[width=\columnwidth]{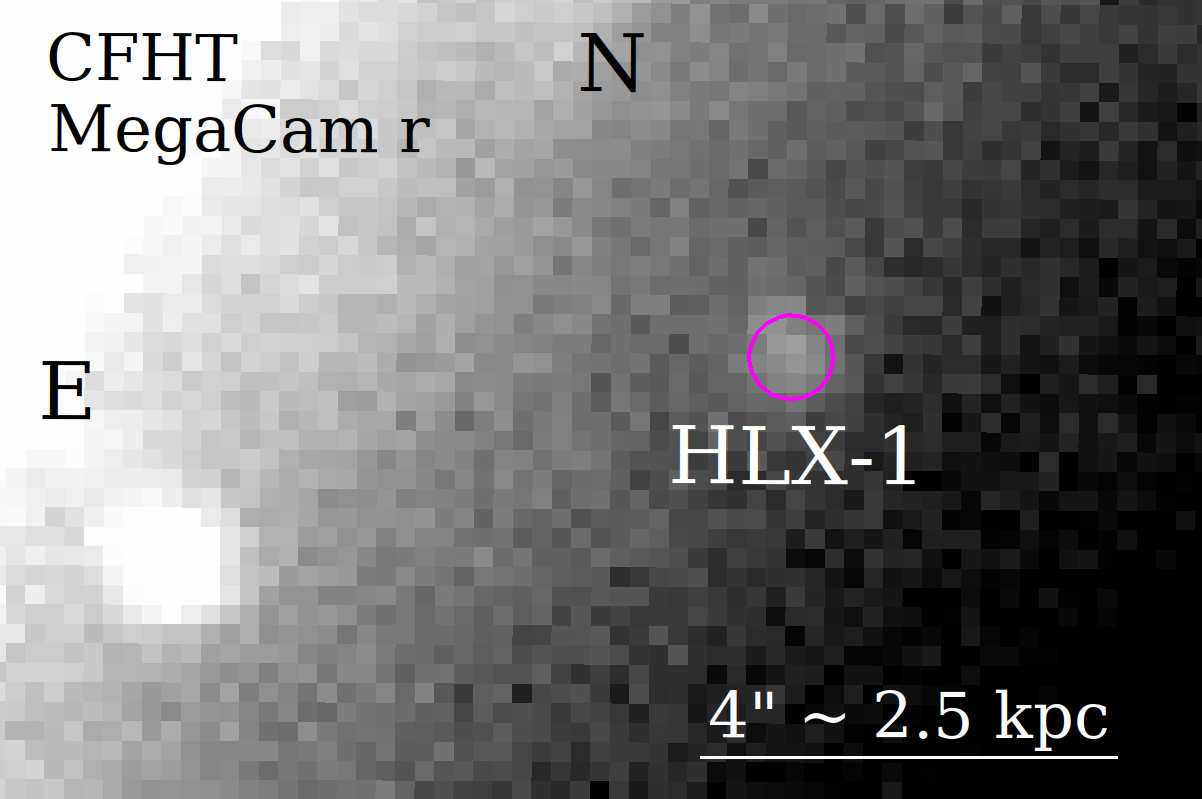}
    \caption{Top panel: CFHT/MegaCam {\it u}-band image (u.MP9302 filter) of the field around HLX-1, taken on 2022 August 2; the 90\% error circle (radius of 0\farcs4) for the position of the corresponding {\it Chandra} source is overplotted in magenta, in all three panels. Middle panel: MegaCam {\it g}-band image (g.MP9402 filter). Bottom panel: MegaCam {\it r}-band image (r.MP9602 filter).
    }
    \label{fig:counterpart}
\end{figure}

\begin{figure}[H]
    \includegraphics[width=\columnwidth]{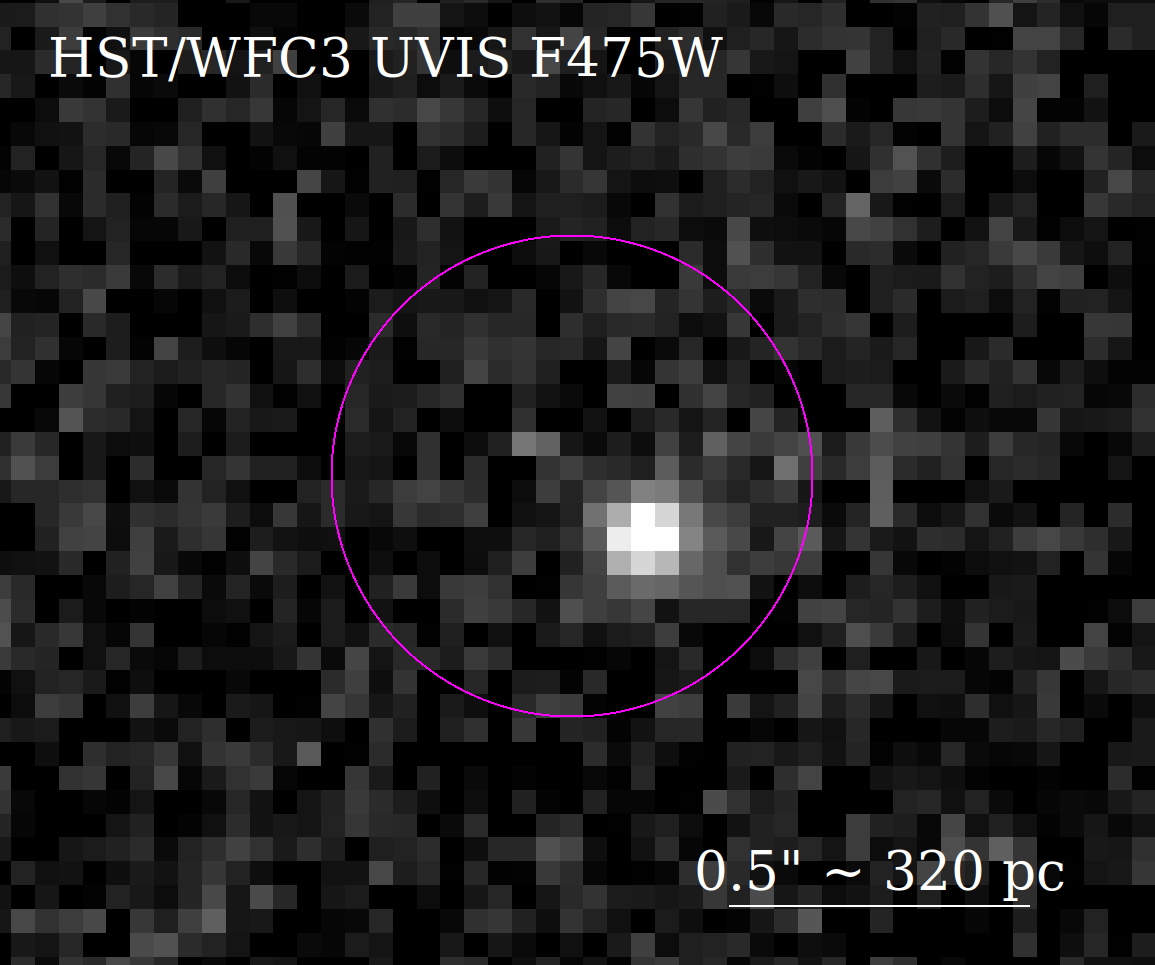}
    \caption{{\it HST}/WFC3 UVIS image in the F475W filter. The green circle is the 90\% uncertainty on the {\it Chandra} source position (radius of 0\farcs4). North is up and east to the left.
    }
    \label{fig:hst_counterpart}
\end{figure}

\section{Optical/X-ray alignment}
\label{section:astrometry}
To search for an optical counterpart of HLX-1, we improved the astrometric solution of the {\it Chandra} and CFHT images, by aligning them onto the same reference frame defined by the {\it Gaia} DR3 catalog \citep{2022yCat}. 
To do so, we searched for {\it Chandra}/{\it Gaia} point-like associations: we found eight such sources. 
We determined the centroid positions of those eight {\it Chandra} sources in the 0.3--8.0 keV band with the {\sc ciao} task {\fontfamily{qcr}\selectfont wavdetect}. 
We then applied the {\fontfamily{qcr}\selectfont wcs\_match} and {\fontfamily{qcr}\selectfont wcs\_update} tasks to the {\it Chandra} data, to align them onto the {\it Gaia} reference frame.
This corresponds to a shift $\Delta$\,R.A. $\approx 0\farcs14$ and $\Delta$\,Dec $\approx 0\farcs44$ with respect to the uncorrected {\it Chandra} astrometry from the data archive. 
After the correction, the residual 1-$\sigma$ random scatter between optical and X-ray positions is $\approx$0\farcs18 in R.A. and $\approx$0\farcs41 in Dec. 
The relatively large residual scatter is partly due to the location of the calibration sources: all eight of them are located farther than 4$^\prime$ from the ACIS-I aimpoint, in regions where the {\it Chandra} point-spread-function is already substantially degraded. 
However, a substantial contribution to the scatter in Dec comes from only one of the eight reference {\it Gaia} sources. 
From a closer analysis of that source (including our inspection of optical images from the Dark Energy Spectroscopic Instrument Legacy Imaging Surveys: \citealt{duncan22}) we infer that the source is partly resolved, with two optical peaks separated by $\approx$0\farcs9 in Dec. 
The {\it Chandra} position appears to correspond to the fainter of the two optical peaks. 
Given the doubts about the accuracy of the X-ray/optical association in this particular case, we decided to remove this calibration source, and use only the other seven. 
This way, the astrometric shift of the {\it Chandra} coordinates with respect to the uncorrected archival data is $\Delta$\,R.A. $\approx 0\farcs12$ and $\Delta$\,Dec $\approx 0\farcs32$, and the residual scatter is $\approx$0\farcs19 in R.A. and $\approx$0\farcs28 in Dec.
\par
We then used {\fontfamily{qcr}\selectfont wavdetect} again to determine the instrumental centroid position of HLX-1 in the astrometrically corrected images, and its 1-$\sigma$ uncertainty (due to the limited number of ACIS counts); the latter is $\sigma_{\rm{RA}} \approx 0\farcs06$, $\sigma_{\rm{Dec}} \approx 0\farcs05$. 
Combining this measurement uncertainty with the uncertainty in the astrometric alignment, we derive a 1-$\sigma$ error in the absolute location of HLX-1 of $\sigma_{\rm{RA}} \approx 0\farcs19$,  $\sigma_{\rm{Dec}} \approx 0\farcs28$. 
In conclusion, the best-fitting position of HLX-1 is R.A.(J2000) $= 16^h\,15^m\,34^{s}.305 (\pm 0^{s}.013)$, Dec.(J2000) $= +19^{\circ}\,27^{\prime}\,08\farcs 06 (\pm 0\farcs28)$.
\par
Next, we improved the astrometric solution of the CFHT images. 
We selected ten point-like reference sources with {\it Gaia} positions.
We used the {\sc iraf} task {\fontfamily{qcr}\selectfont ccfind} to match the reference stars in {\it Gaia} coordinate to the CFHT images.
We then fitted the astrometric solutions (independently for each of the three bands) with the {\sc iraf} task {\fontfamily{qcr}\selectfont ccmap} in interactive mode, removing the reference stars with the highest residuals. 
The best-fit plate solution can then be used to update the image coordinate system via the task {\fontfamily{qcr}\selectfont ccsetwcs} and convert coordinate lists via the task {\fontfamily{qcr}\selectfont cctran}.
Typical shifts of the improved astrometric solutions with respect to the raw images are $\Delta$\,R.A. $\approx 0\farcs07$ and $\Delta$\,Dec $\approx 0\farcs08$ for the $u$-band image, $\Delta$\,R.A. $\approx -0\farcs12$ and $\Delta$\,Dec $\approx -0\farcs09$ for the image of the $g$ band, $\Delta$\,R.A. $\approx 0\farcs08$ and $\Delta$\,Dec $\approx 0\farcs23$ for the $r$-band image.
For the $u$-band image, we obtained a best fit solution with a residual 1-$\sigma$ scattering $\sigma_{\rm{RA}} \approx 0\farcs12$, $\sigma_{\rm{Dec}} \approx 0\farcs10$. 
For the $g$-band and $r$-band images, the astrometric residuals are $\sigma_{\rm{RA}} \approx 0\farcs12$, $\sigma_{\rm{Dec}} \approx 0\farcs10$, consistent with those of the $u$-band image.
\par
Associating the astrometrically corrected {\it Chandra} position of HLX-1 onto the astrometrically corrected CFHT images, we unambiguously identify a blueish optical counterpart (Figure~\ref{fig:fig1}, bottom-right panel). 
The brightness and colours of this source will be discussed in Section~\ref{section:photometry}.
The centroid of the optical source (averaged over the three bands) is R.A.(J2000) $= 16^h\,15^m\,34^{s}.305 (\pm 0^{s}.006)$, Dec.(J2000) $= +19^{\circ}\,27^{\prime}\,08\farcs16 (\pm 0\farcs10)$, where the errors are 1$\sigma$.
This optical position is consistent with the independently determined X-ray position. 
\par
We can further tighten the association of the {\it Chandra} source and its CFHT counterpart with a relative rather than absolute astrometric alignment. 
We identified 11 point-like X-ray sources in the {\it Chandra} image within 5$^\prime$ of the ACIS-I aimpoint, which have an obvious optical counterpart in the CFHT images; only one of these 11 sources is in the {\it Gaia} sample used earlier for absolute astrometric calibration. 
We used these new reference sources to re-align the CFHT image directly onto the {\it Chandra} frame. 
After re-alignment, the 1-$\sigma$ random scatter between X-ray and optical positions is $\approx$0$\farcs$17 in R.A. and in Dec. 
Thus, the 90\% (1.645-$\sigma$) error radius of the {\it Chandra} position over-plotted onto the CFHT images is $\approx$0$\farcs$4 (Figure~\ref{fig:counterpart}).
\par
Finally, we report preliminary results of the {\it HST}/WFC3 observations (K.\ Grishin et al., in prep.). 
The optical counterpart of HLX-1 is detected in all three bands (F300X, F475W and F814W), at a position consistent with the  {\it Chandra} position (Figure~\ref{fig:hst_counterpart}). 
From the {\sc{SExtractor}} \citep{bertin96} catalog associated with the F814W dataset, the centroid of the optical source is R.A.(J2000) $= 16^h\,15^m\,34^{s}.300$, Dec.(J2000) $= +19^{\circ}\,27^{\prime}\,07\farcs94$, with a centroiding 1-$\sigma$ error radius of 0$\farcs$026 (not including the absolute astrometric uncertainty). 
\par
The optical field around HLX-1 is not particularly crowded. There is no other optical source brighter than $m_{g,{\rm Vega}} \approx 25$ mag within 4\arcsec\ of the assumed optical counterpart. 
There are 4 point-like sources brighter than 25 mag within a radius of $\approx$7\arcsec\ (area of $\approx$150 arcsec$^2$). 
The probability that a point-like X-ray source randomly placed in that field ends up at $\lesssim$0\farcs3\ of an optical source is $\approx$1/500. 
The three strongest IMBH candidates in the catalog of \cite{Tranin_2022} (Section~\ref{section:tranin_catalog}) are all located in sparsely populated fields near the outskirts of spheroidal galaxies, and each has a point-like optical counterpart within $\lesssim$0\farcs3\ of its X-ray position.

\begin{table*}
    \centering
    \begin{tabular}{cccccccccc}% ten columns, alignment for each
        \hline
        Band & $t_{\rm exp}$ & $A_\lambda$ & $m_{\rm AB}$ & $m_{\rm AB,0}$ & $m_{\rm Vega,0}$ & $M_{\rm AB}$ & $M_{\rm Vega}$ & $f_{\lambda}$ & $f_{\lambda,0}$ \\
         & (s) & (mag) & (mag) & (mag) & (mag) & (mag) & (mag) & (CGS) & (CGS) \\[4pt]
       \hline
        $u$ & 2800.8 & 0.22 & $24.84 \pm 0.12$ & $24.62 \pm 0.12$ & $24.00 \pm 0.12$ &$-11.10 \pm 0.12$ &$-11.72 \pm 0.12$ & $9.33 \pm 1.03$ & $11.4 \pm 1.3$ \\ [1ex]
        $g$ & 1200.4 & 0.17 & $24.58 \pm 0.16$ & $24.41 \pm 0.16$ & $24.50 \pm 0.16$ & $-11.31 \pm 0.16$ &$-11.22 \pm 0.16$ & $7.03 \pm 1.02$ & $8.22 \pm 1.19$ \\ [1ex]
        $r$ & 2800.8 & 0.11 & $24.95 \pm 0.32$ & $24.84 \pm 0.32$ & $24.66 \pm 0.32$ & $-10.88 \pm 0.32$ & $-11.06 \pm 0.32$ & $2.80 \pm 0.83$ & $3.10 \pm 0.91$ \\ [1ex]
        \hline
    \end{tabular}
    \caption{Brightness of the optical counterpart of HLX-1, measured from the CFHT data. We report both the observed values ($m_{\rm AB}$) and those corrected for a line-of-sight Galactic reddening $E(B-V)= 0.05$ mag ($m_{\rm AB,0}$, $m_{\rm Vega,0}$). The conversion from AB to Vega magnitudes is based on the coefficients listed in the MegaPrime/MegaCam website. Absolute magnitudes $M_{\rm AB}$ and $M_{\rm Vega}$ are dereddened values. Flux densities $f_{\lambda}$ (observed) and $f_{\lambda,0}$ (dereddened) are in units of $10^{-19}$ erg cm $^{-2}$ sec$^{-1}$ \r{A}$^{-1}$.}
    \label{tab:CFHT_info}
\end{table*}

\section{Optical brightness}
\label{section:photometry}
We inspected the radial profile of the optical counterpart of HLX-1 (Section~\ref{section:astrometry}) in the CFHT images with {\sc ds9}. 
In all three bands, the source is consistent with a 2-D Gaussian with a full-width-half-maximum (FWHM) similar to the seeing, {\it i.e.}, between $\approx$0\farcs9 and $\approx$1\farcs3. 
Thus, there is no evidence that the source is extended. 
From the {\it u} band, which has the lowest contamination from the old stellar halo emission of NGC\,6099, we estimate an upper limit of $\approx$0\farcs9 in diameter for the optical counterpart. 
In fact, our preliminary {\it HST} investigation shows (K.~Grishin et al., in prep.) that the counterpart appears point-like even in the WFC3 images, implying an angular size $<$0\farcs15 in diameter ($\approx$100 pc if it is located at the same distance as NGC\,6099).
\par
To measure the optical brightness of HLX-1 in the CFHT images, we selected seven nearby point-like sources with known apparent magnitudes, listed the Sloan Digital Sky Survey (SDSS) Data Release 12 \citep{Alam_2015}. 
We converted the AB brightness values from the SDSS {\it u,g,r} photometric system to the CFHT Megacam {\it u,g,r} system, using the relations in \cite{Gwyn_2008}\footnote{See also \url{https://www.cadc-ccda.hia-iha.nrc-cnrc.gc.ca/en/megapipe/docs/filt.html}.}. 
For the seven reference stars, we adopted source extraction radii of 1\farcs4 for the {\it u} band, 2\farcs2 for the {\it g} band, 2\farcs6 for the {\it r} band; background annuli were taken between 1\farcs9 and 2\farcs6 in the {\it u} band, and between 2\farcs6 and 3\farcs6 for both {\it g} and {\it r} bands.
For HLX-1, we used a source extraction radius of 1\farcs3 in all three CFHT images, and a background annulus between 1\farcs3 and 2\farcs2, to reduce the contamination from the diffuse stellar emission in the halo of NGC\,6099.
We used the {\sc astroimagej} multi-aperture differential photometry tool \citep{Collins_2017} to determine the background-subtracted fluxes from the reference stars and from HLX-1, the latter suitably corrected for its smaller extraction aperture.
\par
For HLX-1, we obtain apparent AB brightnesses $m_{u,{\rm AB}} = (24.84 \pm 0.12)$ mag, $m_{g,{\rm AB}} = (24.58 \pm 0.16)$ mag, $m_{r,{\rm AB}} = (24.95 \pm 0.32)$ mag (Table~\ref{tab:CFHT_info}).
The line-of-sight Galactic reddening \citep{Cardelli_1989, Schlegel_1998} is $E(B-V) \approx 0.05$ mag. 
This corresponds to an extinction $A_{u} \approx 0.22$ mag, $A_{g} \approx 0.17$ mag, $A_{r} \approx 0.11$ mag in the Megacam system \citep{Schlafly_2011}. 
The low value of extinction is due to the high Galactic latitude of HLX-1 ($l = 34\fdg96302$, $b = 42\fdg81152$).
Subtracting the distance modulus of NGC\,6099 ($dm \approx 35.72$ mag), we obtain dereddened absolute brightness $M_{u,{\rm AB}} = (-11.10 \pm 0.12)$ mag, $M_{g,{\rm AB}} = (-11.31 \pm 0.16)$ mag, $M_{r,{\rm AB}} = (-10.88 \pm 0.32)$ mag. 
We also provide (Table~\ref{tab:CFHT_info}) apparent and absolute brightness in the Vega system. 
The conversion from AB to Vega magnitudes is based on the coefficients listed in the MegaPrime/MegaCam website\footnote{\url{https://www.cfht.hawaii.edu/Instruments/Imaging/Megacam/specsinformation.html}.}.
Moreover, we provide an approximate conversion to the Johnson-Cousins UBVR system, using the colour transformations of \cite{Gwyn_2008} between MegaCam and SDSS systems, and then the relations of \citep{jordi06} between SDSS and Johnson-Cousins systems. 
We estimate $U_0 = (23.7 \pm 0.2)$ mag, $B_0 = (24.5 \pm 0.2)$ mag, $V_0 = (24.6 \pm 0.2)$ mag (corrected for line-of-sight reddening). 
\par
Finally, we measured the source brightness from the {\it HST}/WFC3 observations (K.\ Grishin et al., in prep.), in three bands (F300X, F475W and F814W). The {\it HST} images were taken thirteen months after the CFHT images; thus, in principle the source may have varied between the two datasets. 
The apparent brightnesses (AB mag system\footnote{\url{https://www.stsci.edu/hst/instrumentation/wfc3/data-analysis/photometric-calibration/uvis-photometric-calibration}}) is $m_{\rm F300X,AB} = (25.14 \pm 0.05)$ mag in the F300X band (broad UV), $m_{\rm F475W,AB} = (24.62 \pm 0.05)$ mag in the F475W band (similar to the SDSS g$\prime$ band and, roughly, to the $B$ band), and $m_{\rm F814W,AB} = (24.57 \pm 0.05)$ mag in the F814W band (similar to the $I$ band).
All those values were already corrected for line-of-sight Galactic extinction.
Converting to the Vegamag system, and applying a distance modulus of 35.72 mag, those values correspond to extinction-corrected absolute magnitudes $M_{\rm F300X,Vega} = (-11.99 \pm 0.05)$ mag, $M_{\rm F475W,Vega} \approx M_{g\prime,{\rm Vega}} = (-11.00 \pm 0.05)$ mag, $M_{\rm F814W,Vega} \approx M_I = (-11.58 \pm 0.05)$ mag.
In terms of standard (dereddened) colours in the Johnson-Cousins system, putting together CFHT and {\it HST} results, we estimate $U-B = (-0.8 \pm 0.3)$ mag, $B-V =  (-0.1 \pm 0.3)$ mag, $V-I =  (0.5 \pm 0.2)$ mag.

\begin{figure}[t]
    \raggedleft
    \includegraphics[width=1\columnwidth]{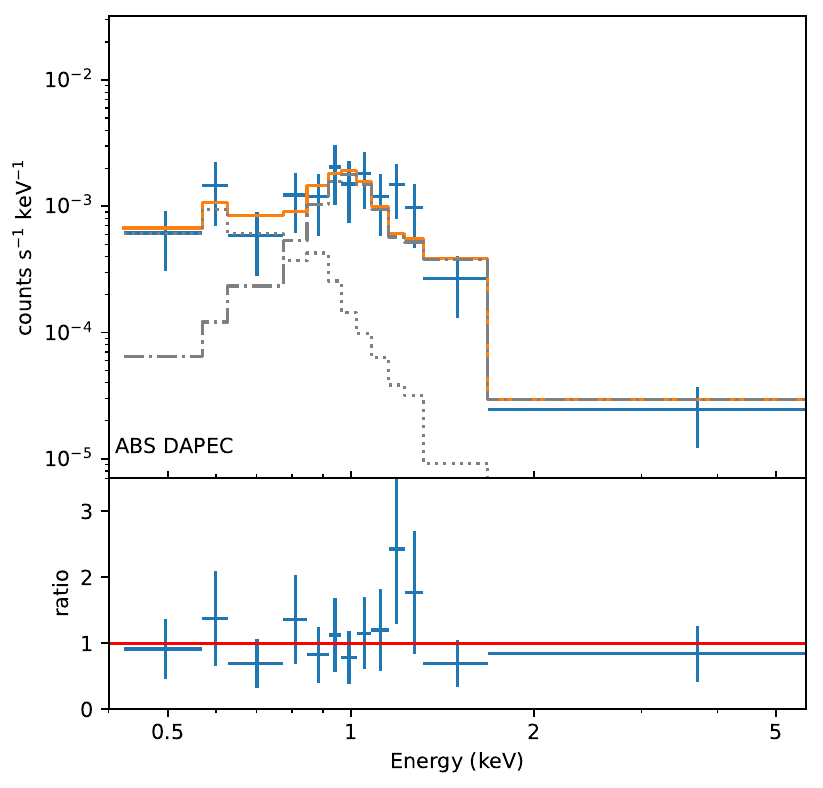}
    \includegraphics[width=0.957\columnwidth]{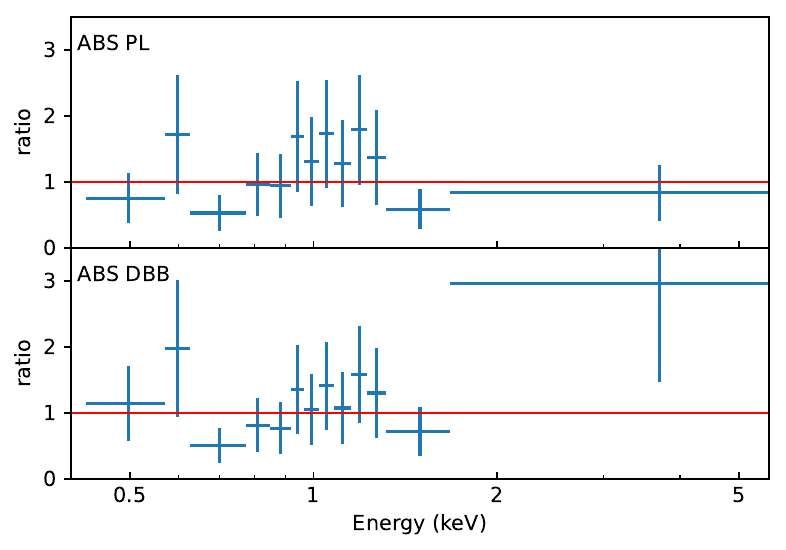}\\
    \caption{Top two panels: 2009 {\it Chandra}/ACIS background-subtracted, folded spectrum, fitted with a double apec model, and corresponding data/model ratios. The data have been fitted with the Cash (Wstat) statistics. They have been subsequently grouped to a minimum signal-to-noise ratio of 1.8 for plotting purposes only.
    See Table~\ref{tab:params_list} for the best-fitting parameters.
    Bottom panels: data/model ratios for the same {\it Chandra}/ACIS spectrum, fitted with alternative models (Table~\ref{tab:params_list}): from top to bottom, a power-law, and a disk-blackbody model.
    }
    \label{fig:chandra_spectrum}
\end{figure}

\section{X-ray Results}
\subsection{Chandra spectrum from 2009}
\label{section:Chandra-obs}
The 2009 {\it Chandra} spectrum of HLX-1 has only $\approx$50 net counts. 
This is enough for simple one-component spectral modelling such as power-law or disk-blackbody (Table~\ref{tab:params_list}). 
In all cases, we fit the spectra with the Cash statistics.
\par
First, we tried a power-law model absorbed only by the line-of-sight Galactic column density. 
We used the Tuebingen-Boulder photoelectric absorption model ({\fontfamily{qcr}\selectfont TBABS} in {\sc xspec}) for this and all other X-ray spectral modelling in this work. 
We applied the {\fontfamily{qcr}\selectfont 'wilm'} abundance table \citep{wilms00} from {\sc xspec}. 
We adopted the Galactic line-of-sight neutral hydrogen column density $N_{\rm H} = 4.16 \times10^{20}$ cm$^{-2}$, obtained from the {\sc colden}\footnote{\url{https://cxc.harvard.edu/toolkit/colden.jsp}.} online tool, based on the radio maps of \citet{Dickey1990}. 
Alternative online tools for the estimation of the hydrogen column density give values ranging from $N_{\rm H} = 3.7 \times10^{20}$ cm$^{-2}$ (HEASARC's {\sc nh} tool\footnote{\url{https://heasarc.gsfc.nasa.gov/cgi-bin/Tools/w3nh/w3nh.pl}.}, based on the \citealt{HI4PI16} map) to $N_{\rm H} = 4.4 \times10^{20}$ cm$^{-2}$ (Swift Science Data Centre's {\sc nhtot} tool\footnote{\url{https://www.swift.ac.uk/analysis/nhtot/docs.php}.}, based on the maps of \citealt{kalberla05}, but including also a contribution from molecular hydrogen, as described in \citealt{Willingale2013}). 
The difference between those alternative choices is negligible for our analysis. 
The photon index of our best-fitting power-law model ({\fontfamily{qcr}\selectfont \emph {ABS PL}} in Table~\ref{tab:params_list}) is $\Gamma = 3.8 \pm 0.6$. 
The absorbed 0.3--10 keV flux is $f_{\rm X} = (2.2_{-0.4}^{+0.6}) \times 10^{-14}$ erg cm$^{-2}$ s$^{-1}$ and the corresponding unabsorbed luminosity is $L_{\rm X} = (8.1_{-2.3}^{+2.1}) \times 10^{40}$ erg s$^{-1}$. 
The fit statistics is Cstat $= 32$ over 47 degrees of freedom (dof). 
To assess whether the inclusion of an additional free intrinsic absorption component significantly improves the model, we created a set of 1000 simulated datasets with the {\fontfamily{qcr}\selectfont {simftest}} script in {\sc{xspec}}; we verified that such additional component is not required.  
\par
The steep value of the photon index for the power-law model suggests that a thermal model is physically more plausible; the latter should also provide a better estimate of the intrinsic luminosity, because it does not diverge at the lower-energy end. 
Therefore, we tried fitting the same spectrum with the standard multicolor blackbody model ({\fontfamily{pcr}\selectfont \emph {diskbb}} in {\sc xspec}).
Such model with only line-of-sight absorption returns a peak color temperature $kT_{\rm in} = (0.22_{-0.04}^{+0.05})$ keV and a normalization corresponding to an apparent inner disk radius $r_{\rm in} = (11,300_{-1,400}^{+1,600}) \,(\cos \theta)^{-1/2}$ km at the distance of NGC\,6099 ({\fontfamily{qcr}\selectfont \emph {ABS DBB}} in Table~\ref{tab:params_list}). 
A ''true'' inner-disk radius $R_{\rm in} \approx 1.19 r_{\rm in}$ \citep{Kubota_1998} is often used in the X-ray modelling of accretion disks, where the correction factor takes into account the no-torque inner boundary condition and the ratio between the color temperature and the effective temperature (hardening factor); the value of 1.19 corresponds to a ''standard'' hardening factor $f \approx 1.7$ \citep{shimura95,merloni00,davis19}. 
With this approximation, the true inner disk radius $R_{\rm in} = (13,400_{-1,600}^{+1,900}) \,(\cos \theta)^{-1/2}$ km.
The absorbed 0.3--10 keV flux is $f_{\rm X} = (1.5_{-1.2}^{+0.2}) \times 10^{-14}$ erg cm$^{-2}$ s$^{-1}$ and the unabsorbed luminosity is $L_{\rm X} =(4.4_{-1.1}^{+1.3}) \times 10^{40}$ erg s$^{-1}$. The fit statistic is Cstat $= 31/47$ dof. 
In this case, too, the addition of a free intrinsic absorption component does not lead to a statistical improvement of the fit. 
The disk-blackbody fit is statistically equivalent to the power-law fit, although we consider the thermal model more physical for an accreting compact object in a soft state. 
More generally, accreting compact objects are typically modelled with thermal plus non-thermal components ({\it e.g.}, disk-blackbody plus power-law) or Comptonization models. 
In the case of our 2009 {\it Chandra}, the number of counts is too low and the systematic uncertainties too high to effectively constrain such multi-component models. 
We verified with a {\fontfamily{qcr}\selectfont {simftest}} analysis that the addition of a second component does not improve the single-component fits.
\par
Another possibility, sometimes usefully applied to ULX spectra \citep{walton18,walton20,barra24}, is a double thermal model ({\it e.g.}, blackbody plus blackbody). 
One component may correspond to the outer (standard) disk emission, the other to the innermost (advective) part of the disk; alternatively, they may be interpreted as emission from the disk and from the photosphere of an optically thick outflow. 
Guided by this possible physical interpretation, we tried fitting a double blackbody model with fixed line-of-sight absorption. 
The fit statistics (Cstat $= 30/45$ dof) is equivalent to those of the power-law and disk-blackbody models; in this case, too, the addition of a free intrinsic absorption component (Cstat $= 29/44$ dof) does not improve the fit.
One thermal component is well constrained, with $kT_1 = (0.18_{-0.03}^{+0.02})$ keV, radius $R_{\rm bb,1} = (16,400 \pm 6,800)$ km and unabsorbed isotropic 0.3--10 keV luminosity $L_{\rm bb,1} = (3.6^{1.3}_{1.0}) \times 10^{40}$ erg s$^{-1}$. The second (cooler) thermal component is unconstrained and its presence does not improve the fit.
\par
Finally, we considered the possibility that the emission is due to an optically thin thermal plasma (for example in the case of shocked gas), and we fitted the 2009 {\it Chandra} spectrum with a double-temperature {\fontfamily{qcr}\selectfont \emph {apec}} model ({\fontfamily{qcr}\selectfont \emph {ABS DAPEC}} in Table~\ref{tab:params_list}, Figure~\ref{fig:chandra_spectrum}) with fixed redshift $z = 0.03$ and fixed solar abundances. 
We obtain good fits (Cstat $=30$/45 dof), for example, with fixed line-of-sight absorption and plasma temperature components $kT_1 = (0.16_{-0.11}^{+0.07})$ keV and $kT_2 = (1.24_{-0.27}^{+0.49})$ keV, corresponding to an unabsorbed 0.3--10 keV luminosity $L_{\rm X} = (6.2_{-1.7}^{+1.6}) \times 10^{40}$ erg s$^{-1}$.
In summary, all the models listed in Table~\ref{tab:params_list} are equally acceptable fits of the 2009 {\it Chandra} spectrum, although they are associated with different physical interpretations.
\par
In addition to the study of HLX-1, we used the {\it Chandra} data to extract and model the spectrum of the diffuse emission around the nucleus and innermost region of NGC\,6099. 
The spectrum is well described (Cstat $= 50$ for 66 dof) by a single-temperature, solar-abundance {\fontfamily{qcr}\selectfont \emph {apec}} model, with fixed redshift ($z = 0.03$), fixed line-of-sight Galactic absorption and best-fitting $N_{\rm H,int} = (5.7_{-0.3}^{+0.4}) \times 10^{21}$ cm$^{-2}$. 
The best-fitting plasma temperature is $kT = (0.37_{-0.16}^{+0.27})$ keV, with an observed 0.3--10 keV flux of $f_{\rm X} = (1.6_{-0.4}^{+0.2}) \times 10^{-14}$ erg cm$^{-2}$ s$^{-1}$. 
In the {\it Chandra} images (Figure~\ref{fig:fig1} bottom-left panel), HLX-1 is sufficiently far away and the ACIS PSF is sufficiently sharp to avoid any contamination from the diffuse galaxy emission. 
However, this is not necessarily the case for the subsequent {\it XMM-Newton} observations. 
By modelling the extent, spectral shape, and flux of the diffuse gas emission from the {\it Chandra} data, we will constrain its contamination to the HLX-1 emission in the {\it XMM-Newton} data.

\begin{table*}
    \caption{X-ray spectral parameters of HLX-1 for selected models. Errors are 90\% confidence limits for one interesting parameter. All models include a line-of-sight absorption $N_{\rm H} = 4.16 \times10^{20}$ cm$^{-2}$, in addition to the intrinsic absorption listed here.}
    \centering
    \resizebox{0.95\textwidth}{!}
    {
    \begin{threeparttable}
    \scriptsize
    \begin{tabular}{*{6}{c}} % six columns, alignment for each
    \toprule
    Observation & Model & Spectral Parameters & Statistic/dof & $F_{\rm X}$ & $L_{\rm X}$ \\
    (1)& (2) & (3) & (4) & (5) & (6) \\
    \midrule
    \multirow{16}{*}{\it Chandra} & \multirow{3}{*}{ABS PL} &  $N_{\rm H,int} = [0]$ & \multirow{3}{*}{32/ 47} & \multirow{3}{*}{$0.22_{-0.04}^{+0.06}$} & \multirow{3}{*}{$0.81_{-0.23}^{+0.21}$} \\
    &&  $\Gamma = 3.8\pm0.6$   &&& \\ 
    && $N_{\rm pl} = 0.43_{-0.10}^{+0.11}$ &&& \\ 
    \cmidrule{2-6}
    & \multirow{3}{*}{ABS DBB} &  $N_{\rm H,int} = [0]$   & \multirow{3}{*}{31/ 47} & \multirow{3}{*}{$0.15_{-0.12}^{+0.02}$} & \multirow{3}{*}{$0.44_{-0.11}^{+0.13}$} \\
    &&   $kT_{\rm in} = 0.22_{-0.04}^{+0.05}$    &&& \\
    && $N_{\rm disk} = 0.66_{-0.13}^{+0.16}$ &&& \\ 
    \cmidrule{2-6}
    & \multirow{5}{*}{ABS DAPEC} &  $N_{\rm H,int} = [0]$ & \multirow{5}{*}{30/ 45} & \multirow{5}{*}{$0.19_{-0.12}^{+0.07}$} & \multirow{5}{*}{$0.62_{-0.17}^{+0.16}$} \\
    &&  $kT_{1} = 0.16_{-0.11}^{+0.07}$  &&& \\
    && $N_{\rm ap1} = 2.3_{-1.5}^{+1.9}$ &&& \\
    && $kT_{2} = 1.24_{-0.27}^{+0.49}$ &&& \\
    && $N_{\rm ap2} = 0.50_{-0.16}^{+0.24}$ &&& \\
    \midrule
    \multirow{11}{*}{XMM1} & \multirow{5}{*}{ABS (PL+DBB)} &  $N_{\rm H,int} = 0.07\pm0.03$ & \multirow{5}{*}{275/ 256} & \multirow{5}{*}{$7.41_{-0.26}^{+0.17}$} & \multirow{5}{*}{$39.3 \pm 5.7$} \\
    &&  $\Gamma = 3.9\pm0.3$  &&& \\
    && $N_{\rm pl} = 15.0_{-3.6}^{+3.8}$ &&& \\
    && $kT_{\rm in} = 0.25_{-0.02}^{+0.03}$ &&& \\
    && $N_{\rm disk} = 7.96_{-3.96}^{+6.56}$ &&& \\
    \cmidrule{2-6}
    & \multirow{6}{*}{ABS (SIMPL$\ast$DBB)} & $N_{\rm H,int} = 0.01_{-0.01}^{+0.02}$ & \multirow{6}{*}{281/ 256} & \multirow{6}{*}{$7.56_{-2.80}^{+0.11}$} & \multirow{6}{*}{$23.1_{-3.4}^{+3.3}$} \\
    && $\Gamma = 4.6_{-0.4}^{+0.2}$ &&& \\
    && FrSc $>0.59$ &&& \\
    && UpSc $ = [0]$ &&& \\
    && $kT_{\rm in} = 0.16\pm0.02$ &&& \\
    && $N_{\rm disk} = 91_{-35}^{+68}$ &&& \\
    \midrule
    \multirow{11}{*}{XMM2} &  \multirow{3}{*}{ABS PL} & $N_{\rm H,int} = 0.09_{-0.05}^{+0.07}$  & \multirow{3}{*}{41/ 49} & \multirow{3}{*}{$0.15 \pm 0.02$} & \multirow{3}{*}{$0.85 \pm 0.15$} \\
    && $\Gamma = 3.7_{-0.5}^{+0.6}$ &&& \\
    && $N_{\rm pl} = 0.51_{-0.08}^{+0.11}$ &&& \\
    \cmidrule{2-6}
    & \multirow{3}{*}{ABS DBB} & $N_{\rm H,int} < 0.02$ & \multirow{3}{*}{53/ 49} & \multirow{3}{*}{$0.12_{-0.01}^{+0.02}$} & \multirow{3}{*}{$0.37_{-0.07}^{+0.06}$} \\
    && $kT_{\rm in} = 0.24 \pm 0.03$ &&& \\
    && $N_{\rm disk} = 0.33_{-0.12}^{+0.20}$ &&& \\
    \cmidrule{2-6}
    & \multirow{5}{*}{ABS (PL+DBB)} & $N_{\rm H,int} = 0.19_{-0.12}^{+0.08}$ & \multirow{5}{*}{37/ 47} & \multirow{5}{*}{$0.14_{-0.03}^{+0.04}$} & \multirow{5}{*}{$1.1 \pm 0.3$} \\
    && $\Gamma = 3.6_{-0.8}^{+0.4}$ &&& \\
    && $N_{\rm pl} = 0.55_{-0.20}^{+0.07}$ &&& \\
    && $kT_{\rm in} = 0.07 \pm 0.01$ &&& \\
    && $N_{\rm disk} = 850_{-725}^{+856}$ &&& \\
    \bottomrule
    \end{tabular}
    \begin{tablenotes}
    \scriptsize
%    \item Notes: 
    \par
    Col.(1): see Table~\ref{tab:obs_info} for the observation log
    \par
    Col.(2): the short-hand notation for model components is ABS = {\fontfamily{qcr}\selectfont \emph{Tbabs}}, PL = power-law, DBB = disk-blackbody, DAPEC = two-temperature {\fontfamily{pcr}\selectfont \emph {apec}}, SIMPL = Comptonization model.
    In addition, the {\it XMM-Newton} spectra include a contamination component from the diffuse gas emission of NGC\,6099, modelled as a fraction $C$ of the total emission modelled from the {\it Chandra} data, namely $C \times$  {\fontfamily{qcr}\selectfont \emph{Tbabs}$\times$\emph{Tbabs}$_{int}\times$\emph{apec}}. For XMM1, $C \equiv 0.06$; for XMM2, $C \equiv 0.04$.
    \par
    Col.(3): $N_{\rm H,int}$ is in units of $10^{22}$ cm$^{-2}$; $N_{\rm pl}$ in units of $10^{-5}$ photons keV$^{-1}$ cm$^{-2}$ s$^{-1}$ at 1 keV; $kT_{\rm in}$, $kT_{1}$, $kT_{2}$ in units of keV; $N_{\rm disk} \equiv (r_{\rm in}/D_{10})^{2}\cos\theta$, where the $r_{\rm in}$ is the apparent inner disk radius in km and $D_{10}$ the distance in units of 10 kpc; $N_{\rm ap1}$, $N_{\rm ap2}$ are $10^{-5}$ times the {\sc xspec} {\fontfamily{pcr}\selectfont \emph {apec}}-model normalization explained at {\url {https://heasarc.gsfc.nasa.gov/xanadu/xspec/manual/node134.html}}.
    \par
    Col.(4): the fit statistics is C-stat for {\it Chandra} and $\chi^2$ for XMM1, XMM2. 
    \par
    Col.(5): 0.3--10 keV model absorbed flux, in units of 10$^{-13}$ erg cm$^{-2}$ s$^{-1}$, and 90\% confidence interval.
    \par
    Col.(6): 0.3--10 keV unabsorbed isotropic luminosity, in units of 10$^{41}$ erg s$^{-1}$, and 90\% confidence interval. 
    \end{tablenotes}
    \end{threeparttable}
    }
    \label{tab:params_list}
\end{table*}

\begin{figure}
	% Allowable file formats are eps or ps if compiling %using latex or pdf, png, jpg if compiling using pdflatex
%    \centering
    \raggedleft
    \includegraphics[width=1\columnwidth]{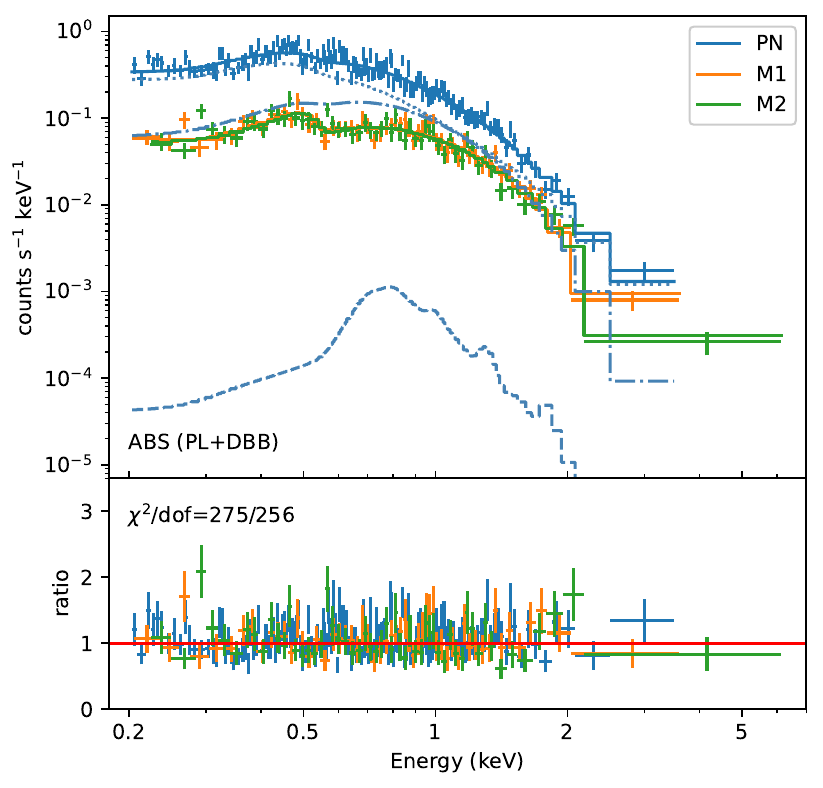}\\
    \includegraphics[width=0.955\columnwidth]{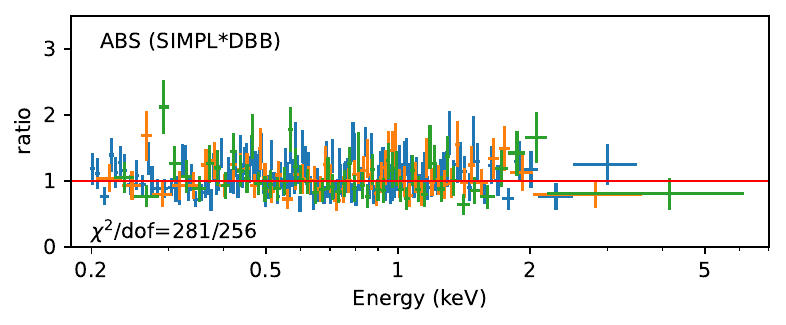}\\   
    \caption{Top two panels: 2012 {\it XMM-Newton}/EPIC background-subtracted, folded spectra (blue: pn; orange: MOS1; green: MOS2), simultaneously fitted with a disk-blackbody plus power-law model, and corresponding data/model ratios. The dotted line shows the power-law contribution for the pn, and the dash-dotted line is the disk-blackbody contribution. The dashed line represents our estimated (fixed) contamination from the diffuse gas in NGC\,6099; its inclusion and assumed shape does not significantly affect any of the the best-fitting parameters. The data have been binned to a minimum of 20 counts per bin, for $\chi^2$ fitting. 
    See Table~\ref{tab:params_list} for the best-fitting parameters (XMM1 observation, {\fontfamily{pcr}\selectfont \emph {ABS (PL+DBB)}} model).
    Bottom panel: data/model ratios for the same {\it XMM-Newton}/EPIC spectra, fitted with Comptonization model ({\fontfamily{pcr}\selectfont \emph {ABS (SIMPL$\ast$DBB)}} in Table~\ref{tab:params_list}).
    }
    \label{fig:xmm-1_spectrum}
\end{figure}

\subsection{XMM-Newton spectra from 2012}
\label{section:XMM1_obs}
In the 2012 {\it XMM-Newton} observations, the flux from HLX-1 dramatically increased by a factor of about 50 compared with 2009, to the point that it was now completely dominant over the diffuse galactic emission.
Generally speaking, we need to carefully assess cross-contamination in a situation where the distance between HLX-1 and the nucleus of NGC\,6099 is comparable to the half-energy width of the on-axis PSF of MOS1, MOS2 and pn \citep{S.L.O'Dell2010}. 
To quantify the amount of residual contamination, we estimated what fraction of the diffuse galactic emission falls within the 14\farcs5 source extraction circle of HLX-1. 
For this, we took the spatial extent of the diffuse emission from the {\it Chandra} observations and convolved it with the EPIC-MOS and EPIC-pn PSFs. 
We estimate that $\approx$6\% of the diffuse gas emission falls inside the HLX-1 source region. Therefore, we include this component by rescaling the normalization of the best-fitting thermal plasma model for the 2009 {\it Chandra} data (Section~\ref{section:Chandra-obs}). 
In any case, the galactic contamination contributes an insignificant flux of $\approx 10^{-15}$ erg cm$^{-2}$ s$^{-1}$ to the 2012 spectra, compared with the observed source flux of $\approx 7 \times 10^{-13}$ erg cm$^{-2}$ s$^{-1}$.
\par
The 2012 dataset is the only one where we have enough signal-to-noise to attempt also a timing analysis. 
We searched for periodicities or quasi-periodic oscillations with standard tasks in the {\sc{ftools}} package ({\fontfamily{qcr}\selectfont {efsearch}} and {\fontfamily{qcr}\selectfont {powspec}}, respectively), in the 1--10,000\,s range, but did not find any. 
Following standard usage \citep[{\it e.g.},][]{vaughan03,heil12,sutton13,middleton15,robba21}, we measured the normalized excess variance in the pn light curve (binned to 100 s) to quantify the short-term flux variability, with the {\sc{ftools}} task {\fontfamily{qcr}\selectfont {lcstats}}. 
In particular, the fractional root mean square (rms) variability amplitude is the square root of the normalized excess variance. 
We estimate an rms of ($20\pm2$) percent. 
We leave further investigations on the X-ray timing properties to follow-up work; here we focus instead on the spectral properties. 
\par
For our spectral analysis, we fitted the pn, MOS1 and MOS2 data simultaneously, with the $\chi^2$ statistics. 
We tested two simple phenomenological models: a power-law and a disk black body, both with fixed line-of-sight absorption plus free intrinsic absorption.
For the absorbed power-law fit, we obtained $\chi^2 = 312$ (258 dof), while for the absorbed disk-blackbody fit, $\chi^2 = 375$ (258 dof). Both models fail to describe the data and are rejected at the $2.5\sigma$ and $4.7\sigma$ level, respectively.  
\par
Significantly better results are obtained with an absorbed {\fontfamily{qcr}\selectfont \emph {PL}} plus {\fontfamily{qcr}\selectfont \emph {DBB}} model (Table~\ref{tab:params_list} for the fit parameters; upper panel of Figure~\ref{fig:xmm-1_spectrum} for a plot of the model fit and residuals): with a statistic of $\chi^{2} = 275$ (256 dof), the model is acceptable within the 90\% confidence level.
The thermal disk component has a peak colour temperature $kT_{\rm in} = 0.25_{-0.02}^{+0.03}$ keV and an apparent inner-disk radius $r_{\rm in} = (39,200^{+16,400}_{-10,200}) \,(\cos \theta)^{-1/2}$ km ({\it{i.e.}}, $R_{\rm in} = (46,700^{+19,500}_{-12,100}) \,(\cos \theta)^{-1/2}$ km). 
The power-law component has a steep photon index $\Gamma = 3.9\pm0.3$. The unabsorbed luminosity in the 0.3--10 keV band is $L_{\rm X} \approx (3.9 \pm 0.6) \times 10^{42}$ erg s$^{-1}$.
\par
We then explored another fitting model often used for soft-state ULXs. We replaced the standard disk with a $p$-free disk ({\fontfamily{qcr}\selectfont \emph {diskpbb}} in {\sc xspec}), in which the temperature scales as $T \propto R^{-p}$ instead of $T \propto R^{-0.75}$ \citep{mineshige94,watarai00,kubota04,kubota05}. This ''broadened disk'' is a well-known approximation of an advection-dominated, super-critical slim disk, with a reduced radiative efficiency in the inner region. Typical values of $p \approx 0.6$ are found in super-critical disks \citep{vierdayanti08,sutton17}.
In our case, we do find a best-fitting value $p \approx 0.6$ but the whole range $0.5 \lesssim p \lesssim 1.0$ is acceptable within the 90\% confidence limit. 
Thus, the model is indistinguishable from a standard disk blackbody with $p=0.75$. 
The goodness-of-fit parameter $\chi^2  = 275$ (255 dof) is unchanged. 
The main reason why we cannot distinguish between power-law plus standard disk and power-law plus $p$-free disk is that most of the emission is in the power-law component.
\par
Finally, we tried Comptonization models, as a more physical representation of the thermal plus non-thermal spectrum. 
We found better fits with models based on multi-temperature thermal seed components, such as {\fontfamily{qcr}\selectfont \emph {diskir}} \citep{gierlinski08,gierlinski09} and {\fontfamily{qcr}\selectfont \emph {simpl}} \citep{Steiner_2009}, rather than single-temperature blackbody seeds such as {\fontfamily{qcr}\selectfont \emph {comptt}} and {\fontfamily{qcr}\selectfont \emph {bmc}}. 
For the ${\fontfamily{pcr}\selectfont \emph{simpl}} \ast{\fontfamily{pcr}\selectfont \emph{diskbb}}$ model, we confirm a very steep power-law slope ($\Gamma = 4.6_{-0.4}^{+0.2}$), with a slightly lower peak temperature ($kT_{\rm in} = 0.16\pm0.02$ keV) compared with that found in the power-law plus disk-blackbody model. 
The apparent inner radius is $r_{\rm in} = (132,000^{+43,000}_{-28,000}) \,(\cos \theta)^{-1/2}$ km ({\it{i.e.}}, $R_{\rm in} = (158,000^{+51,000}_{-34,000}) \,(\cos \theta)^{-1/2}$ km).({\fontfamily{pcr}\selectfont \emph {ABS (SIMPL$\ast$DBB)}} in Table~\ref{tab:params_list}.)
The unabsorbed 0.3--10 keV luminosity is $L_{\rm X} = (2.3 \pm 0.3) \times 10^{42}$ erg s$^{-1}$. Essentially identical parameters are obtained with {\fontfamily{qcr}\selectfont \emph {diskir}}: $\Gamma = 4.3_{-0.3}^{+0.4}$, $kT_{\rm in} = 0.16_{-0.03}^{+0.04}$ keV, $L_{\rm X} = (2.3 \pm 0.3) \times 10^{42}$ erg s$^{-1}$. 
In summary, Comptonization models (${\fontfamily{pcr}\selectfont \emph{simpl}}$ model: $\chi^2 = 281/256$ dof; ${\fontfamily{pcr}\selectfont \emph {diskir}}$ model: $\chi^2 = 283/253$ dof) do not provide better fits than the disk-blackbody plus power-law model but they are still acceptable at the 90\% confidence level. 
Additionally, unabsorbed luminosities estimated from Comptonization models are more reliable, because they avoid the unphysical divergence of the steep power-law at the low-energy end.

\subsection{XMM-Newton spectra from 2023}
\label{section:XMM2_obs}
In 2023, the observed X-ray flux of HLX-1 was back at a level similar to that of 2009 (Table~\ref{tab:params_list}). 
To account for contamination from the NGC\,6099 diffuse emission, we followed the procedure outlined in Section~\ref{section:XMM1_obs}.
Since the X-ray flux of HLX-1 in 2023 was significantly lower than in 2012, optimizing the signal-to-noise ratio required the use of a smaller source region with a radius of 10$\arcsec$. 
We estimate that this region includes approximately 4\% of the total diffuse emission from the galaxy, and we have accordingly rescaled the normalization to account for this leakage.
The presence of this galaxy component is implicit in the fit results of Table~\ref{tab:params_list}, where we only list the fit parameters for the HLX-1 emission. 

\begin{figure}
    \raggedleft
    \includegraphics[width=\columnwidth]{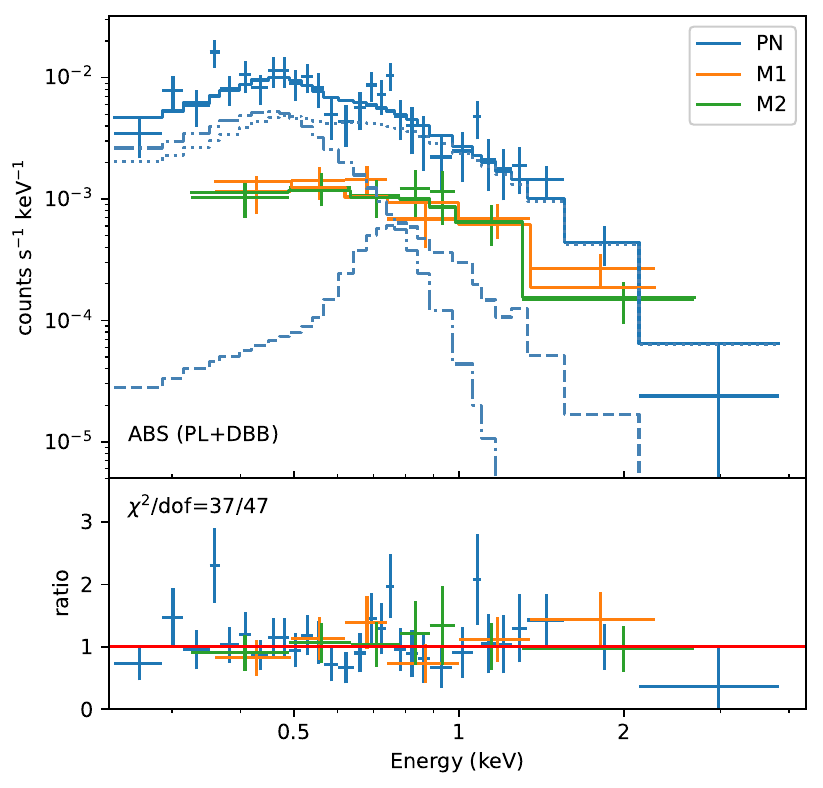}\\
    \includegraphics[width=0.955\columnwidth]{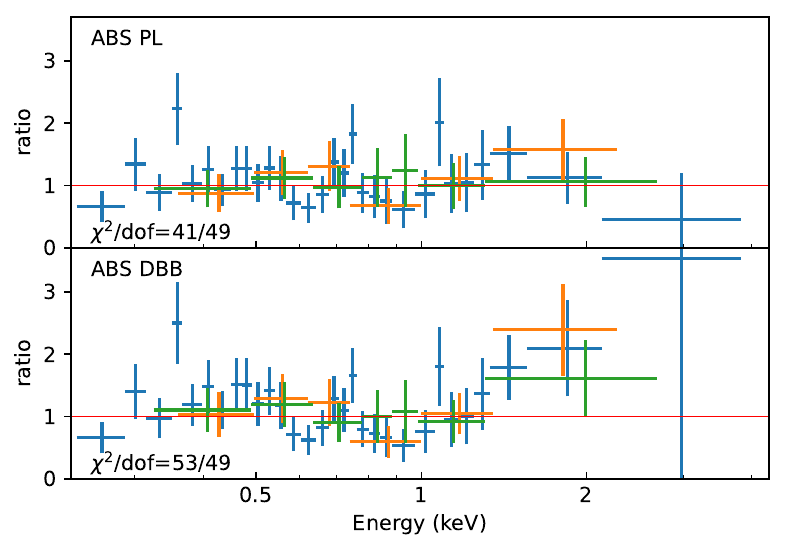}\\
    \caption{Top two panels: 2023 {\it XMM-Newton}/EPIC background-subtracted, folded spectra (blue: pn; orange: MOS1; green: MOS2), simultaneously fitted with a disk-blackbody plus power-law model, and corresponding data/model ratios. The dotted line shows the power-law contribution for the pn, and the dash-dotted line is the disk-blackbody contribution. The dashed line represents our estimated (fixed) contamination from the diffuse gas in NGC\,6099. The data have been binned to a minimum of 20 counts per bin, for $\chi^2$ fitting. 
    See Table~\ref{tab:params_list} for the best-fitting parameters (XMM2 observation, {\fontfamily{pcr}\selectfont \emph {ABS (PL+DBB)}} model).
    Bottom two panels: data/model ratios for the same {\it XMM-Newton}/EPIC spectra, fitted with two alternative models: a power-law model, and a disk blackbody model.
    }
    \label{fig:XMM2_spectrum}
\end{figure}

\begin{figure}
    \includegraphics[width=\columnwidth]{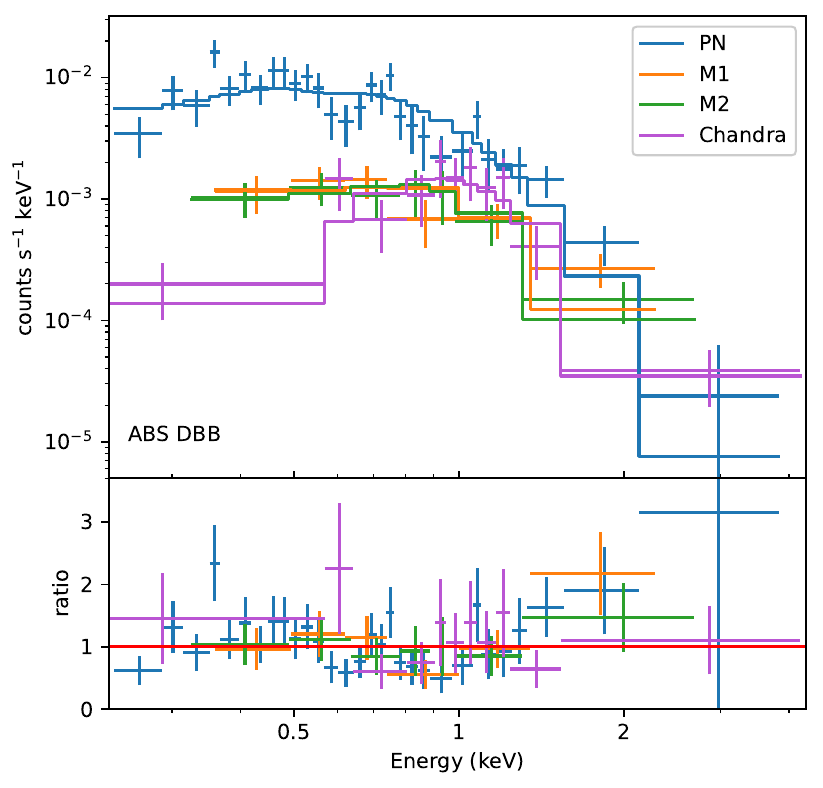}\\
    \caption{Simultaneous fit of the 2009 {\it Chandra} spectrum (orchid datapoints and model fit) and 2023 {\it XMM-Newton}/EPIC spectrum (blue = pn, orange = MOS1, green = MOS2) with an absorbed disk-blackbody model. Top panel: datapoints and best-fitting model; bottom panel: data/model ratios. For all four spectra, the data have been grouped to a minimum signal-to-noise ratio of 2, for plotting purposes only.
    }
    \label{fig:XMM2_chandra_diskbb}
\end{figure}

\begin{figure}
\includegraphics[width=\columnwidth]{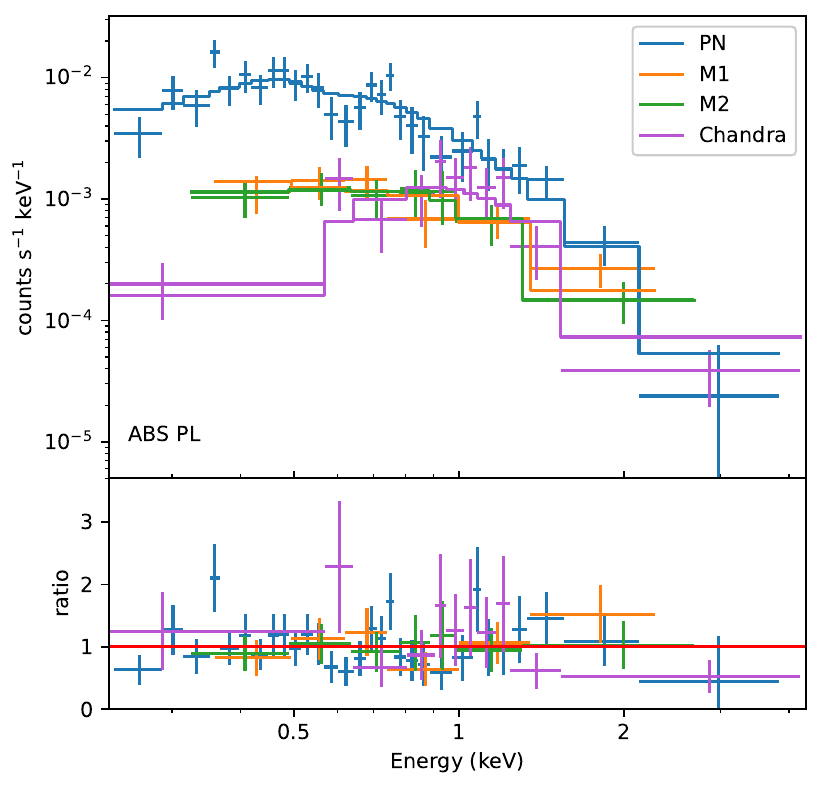}\\
    \caption{As in Figure \ref{fig:XMM2_chandra_diskbb}, for an absorbed power-law model.
    }
    \label{fig:XMM2_chandra_pow}
\end{figure}
\par
We started again from simple one-component models (power law and disk-blackbody) and then a combined power law plus disk-blackbody model. 
We tested the significance of additional free intrinsic absorption with the likelihood ratio test {\fontfamily{qcr}\selectfont {simftest}}. 
For the disk-blackbody model, intrinsic absorption is not required. 
For models that include a power law, intrinsic absorption is required at the 99\% confidence level.  

All three models (power law, disk-blackbody and power law plus disk-blackbody) give acceptable fits (Table~\ref{tab:params_list}). 
The power-law model has a steep slope $\Gamma = 3.7_{-0.5}^{+0.6}$ (Table~\ref{tab:params_list}), and the addition of intrinsic absorption avoids a divergence from the observed flux at the low-energy end. 
When a disk-blackbody component ($kT_{\rm in} = (0.07\pm0.01)$ keV) is also added, there is an inevitable degeneracy between the strength of the soft thermal component and the amount of intrinsic absorption. 
The power law plus disk-blackbody model is generally preferred for physical reasons, but it may over-estimate the unabsorbed luminosity. 
Instead, a pure disk-blackbody model with $kT_{\rm in} = (0.24\pm0.03)$ keV provides a better estimate of the intrinsic luminosity, because of its more physical turnover at low energies. 

For a pure disk-blackbody model, the apparent inner disk radius $r_{\rm in} = (8,000^{+2,100}_{-1,600}) \,(\cos \theta)^{-1/2}$ km ($R_{\rm in} = (9,500^{+2,500}_{-1,900}) \,(\cos \theta)^{-1/2}$ km), marginally consistent with the characteristic emission radius from the 2009 observation. 
Instead, for the power-law plus disk-blackbody model, the thermal component is only a non-dominant correction to the power-law emission; in that case, the radius of the thermal component is poorly constrained and degenerate with the intrinsic absorption.
The unabsorbed, isotropic 0.3--10 keV luminosity $L_{\rm X} = (3.7_{-0.7}^{+0.6}) \times 10^{40}$ erg s$^{-1}$ for the disk-blackbody fit, while $L_{\rm X} \approx 10^{41}$ erg s$^{-1}$ for models dominated by the steep power-law. 

\par
Finally, we tried two Comptonization models: {\fontfamily{pcr}\selectfont \emph {diskir}} and {\fontfamily{pcr}\selectfont \emph {simpl}}.  
In both cases, we obtain statistically equivalent fits to the simple power-law model ($\chi^2 = 37$ over 45 dof for {\fontfamily{pcr}\selectfont \emph {diskir}}, $\chi^2 = 37$ over 47 dof for {\fontfamily{pcr}\selectfont \emph {simpl}}, $\chi^2 = 41$ over 49 dof for a simple power law), because the signal-to-noise ratio and the number of datapoints are too low to require complex multi-component fitting.  
At the high-energy end, because of the steep spectral slope ($\Gamma > 3$), it is not possible to determine whether the spectrum steepens even further (break or exponential cutoff) above a few keV, as is usually the case for super-Eddington stellar-mass ULXs \citep{walton18,walton18b}. 
At the low-energy end, the temperature of the seed thermal component is again degenerate with the amount of intrinsic absorption.
For the {\fontfamily{pcr}\selectfont \emph {diskir}} model, we obtained an intrinsic column density $N_{\rm H,int} = (1.2^{+0.7}_{-0.6}) \times 10^{21}$ cm$^{-2}$, peak color temperature $kT_{\rm in} = (0.08^{+0.01}_{-0.03})$ keV, photon index $\Gamma = 3.4_{-0.4}^{+0.7}$ (Table~\ref{tab:diskir}). 
The apparent inner disk radius is $r_{\rm in} = (164,000^{+167,000}_{-50,000}) \,(\cos \theta)^{-1/2}$ km ($R_{\rm in} = (195,000^{+99,000}_{-60,000}) \,(\cos \theta)^{-1/2}$ km).
The unabsorbed luminosity is $L_{\rm X} = (1.0 \pm 0.2) \times 10^{41}$ erg s$^{-1}$. 
For the {\fontfamily{pcr}\selectfont \emph {simpl} $\times$ \emph {diskbb}} model, $N_{\rm H,int}=(1.7_{-1.0}^{+0.5}) \times 10^{21}$ cm$^{-2}$, $kT_{\rm in} = (0.07_{-0.04}^{+0.01})$ keV, and $\Gamma = 3.6_{-0.3}^{+0.9}$ (Table~\ref{tab:bbodyrad}). 
The inner disk radius is $r_{\rm in} = (370,000^{+223,000}_{-139,000}) \,(\cos \theta)^{-1/2}$ km ($R_{\rm in} = (440,000^{+265,000}_{-165,000}) \,(\cos \theta)^{-1/2}$ km).
The unabsorbed luminosity is $L_{\rm X} \approx (1.6 \pm 0.3) \times 10^{41}$ erg s$^{-1}$.
An important caveat of both Comptonization models is that the best-fitting radius of the thermal component is much larger than the radius obtained from a disk-blackbody model. 
Physically, such radii may represent the size of the Comptonizing region rather than the innermost stable orbit.
We will use Comptonization models again in Section~\ref{section:SED}, to match the UV/optical data to the soft X-ray data, because such models avoid an unphysical low-energy divergence of the power-law component.

\subsection{X-ray comparison between 2009 and 2023}
\label{section:combinefit}
The 2009 {\it Chandra} and 2023 {\it XMM-Newton} spectra have similar fluxes and luminosities (Table~\ref{tab:params_list}), within the uncertainties of the various models. 
It is natural to test whether HLX-1 had ``returned'' to the same X-ray state it was in 2009, after an outburst in between those years. 
To test this possibility, we tried fitting simultaneously the 2009 and 2023 spectra, locking all parameters. 
A short discussion of the disk-blackbody model fit will suffice to illustrate the result. 
We find (Figure~\ref{fig:XMM2_chandra_diskbb}) a perhaps surprisingly good fit, with Cstat $= 93$ (31 $+$ 62 from the 2009 and 2023 datapoints, respectively) over 98 dof. 
The intrinsic absorption column density is negligible ($N_{\rm H,int} < 2 \times 10^{20}$ cm$^{-2}$). 
The best-fitting peak color temperature is $kT_{\rm in} = (0.24 \pm 0.02)$ keV. 
The disk normalization $N_{\rm disk} = 0.36^{+0.13}_{-0.10}$ km$^2$ corresponds to an apparent inner radius $r_{\rm in} = (8,300^{+1,600}_{-1,300}) \,(\cos \theta)^{-1/2}$ km.  
We find an even better agreement with a power-law model (Figure~\ref{fig:XMM2_chandra_pow}), with Cstat $= 76$ (32 $+$ 44 from the 2009 and 2023 datapoints, respectively) over 98 dof. 
In this case, $N_{\rm H,int} = 1.0^{+0.7}_{-0.5} \times 10^{21}$ cm$^{-2}$, and $\Gamma = 3.83^{+0.54}_{-0.46}$. 
In summary, we cannot rule out the possibility that HLX-1 was in the same spectral state in 2009 and 2023.

\section{Origin of the optical emission}
\label{section:SED}
The point-like optical/UV counterpart (Section~\ref{section:photometry}) is too bright ($M_{V} \approx -11$ mag) to be an individual star.
Given its relatively blue optical colours, it could be a young, massive star cluster in the halo of NGC\,6099 (a very atypical situation for an elliptical galaxy); or it could be dominated by reprocessed emission from the X-ray source; or it could be a background AGN/quasar. 
More complicated scenarios are also possible, such as blue/UV emission from X-ray reprocessing in the accretion flow, and red/IR emission from a host old star cluster.
This uncertainty is reminiscent of the debate about the interpretation of the optical counterpart of the IMBH candidate ESO\,243-49 HLX-1, initially interpreted as a young star cluster \citep{Farrell_2012} and then attributed (at least for the blue/UV component) to reprocessed disk emission \citep{Soria_2012,soria17} because its brightness varied together with the X-ray flux. 
In the case of HLX-1, we do not have a spectroscopic optical redshift, to test the background AGN scenario, nor do we have multiple-epoch observations to test the correlation with X-ray luminosity. 
For now, we can constrain the X-ray/optical flux ratio to determine which of the three alternatives (IMBH disk, star cluster, and AGN) is more likely. 
We do not have contemporaneous optical and X-ray observations; our CFHT data were taken in 2022 August, the {\it HST} data in 2023 September, and the closest X-ray observation was from 2023 August.
There is only marginal evidence of a small optical variability between 2022 and 2023, based on the comparison between the extinction-corrected CFHT measurement of $m_{g,{\rm AB}} = (24.41 \pm 0.16)$ mag and the {\it HST} value of $m_{\rm F475W,AB} \approx g' \approx g = (24.62 \pm 0.05)$ mag (Section~\ref{section:photometry}). 
In any case, the change is small enough that we can use both the 2022 and 2023 optical data together, for a comparison with the 2023 X-ray data. 
\par
To model the broadband spectral energy distribution, we combined the 2023 {\it XMM-Newton}/EPIC measurements with six optical/UV datapoints ($u,g,r$ from CFHT/Megacam; F300X, F475W and F814W from {\it HST}/WFC3). 
We used the {\sc ftools} task {\fontfamily{qcr}\selectfont ftflx2xsp} to convert each of the six optical/UV flux measurements into {\it pha} format suitable for {\sc xspec} modelling. 
We tried two models: one in which the optical/UV datapoints are tied to the soft X-ray emission ({\fontfamily{pcr}\selectfont \emph {diskir}}) and one in which X-ray and optical bands are uncorrelated; in the latter case, we used the {\fontfamily{pcr}\selectfont \emph {simpl}} Comptonization model for the X-ray emission and a blackbody model for the optical datapoints.

\begin{table}
    \caption{Optical/X-ray spectral energy distribution from the 2022--2023 data, fitted with an irradiated disk model: 
    \(\fontfamily{pcr}\selectfont \emph {redden} \times \emph{redden}_{int} \times \emph{TBabs} \times (\emph{TBabs}_{int} \times \emph{diskir}  \left[+ 0.04 \times \emph{TBabs}_{\rm gal} \times \emph{apec}_{\rm gal}\,\right])\).
    The last term accounts for the contamination of thermal emission from the host galaxy NGC\,6099; we did not include this term in the flux and luminosity of HLX-1. Errors are 90\% confidence limits for one interesting parameter. The X-ray line-of-sight absorption is $N_{\rm H} \equiv 4.16\times 10^{20}$ cm$^{-2}$; the line-of-sight reddening is $E(B-V) \equiv 0.05$ mag.
    }
    \centering
    {
    \begin{tabular}{lccc} % three columns, alignment for each
        \hline
        Component & Parameter & Value & Unit\\
        \hline
        \(\fontfamily{pcr}\selectfont \emph{redden}_{int}\) & $E(B-V)$  & $0.11_{-0.07}^{+0.21}$  & mag  \\ [1ex]
        \(\fontfamily{pcr}\selectfont \emph{TBabs}_{int}\) & $N_{\rm H, int}$ & $0.12_{-0.06}^{+0.07}$ & $10^{22}$ cm$^{-2}$ \\ [1ex]
        \multirow{9}{*}{\fontfamily{pcr}\selectfont \emph{diskir}} & $kT_{\rm in}$ &  $0.08_{-0.03}^{+0.01}$ &  keV \\ [1ex]
         & $\Gamma$  & $3.4_{-0.4}^{+0.7}$ & \\ [1ex]
         & $kT_{\rm e}$ & [100]  & keV  \\ [1ex]
         & $L_{c}/L_{d}$  & $0.22_{-0.12}^{+1.08} $ & \\ [1ex]
         & $f_{\rm in}$ & [0.1] & \\ [1ex]
         & $r_{\rm irr}$ & [1.2] & \\ [1ex]
         & $f_{\rm out}$ & $0.03_{-0.01}^{+0.07}$ & \\ [1ex]
         & $\log(R_{\rm out})^{a}$ & $3.49_{-0.42}^{+0.21}$ & See Notes\\ [1ex]
         & $N_{\rm disk}^{b}$ &  $140_{-73}^{+427} $ & km$^{2}$ \\ [1ex]
        \(\fontfamily{pcr}\selectfont \emph{TBabs}_{\rm gal}\) & $N_{\rm H,gal}$ & [0.57] & $10^{22}$ cm$^{-2}$  \\ [1ex]
        \multirow{2}{*}{$\fontfamily{pcr}\selectfont \emph {apec}_{\rm gal}$} & $kT$ & [0.37] & keV\\ [1ex]
         & $N_{\rm apec}^{c}$ & [6.85] & See Notes\\ [1ex]
        \hline
         & $\chi^2$/dof & 41/50 & \\ [1ex]
        \hline 
         & \(R_{\rm in} \sqrt{\cos\theta}\) & $19.5_{-5.3}^{+29.9}$  & $10^{9}$ cm\\ [1ex]
         & $F_{\rm X}^{d}$ &  $1.4_{-0.3}^{+0.1} \times 10^{-14}$ & erg cm$^{-2}$ s$^{-1}$ \\ [1ex]
         & $L_{\rm X}^{e}$  & $1.1_{-0.2}^{+0.2} \times 10^{41}$ & erg s$^{-1}$\\ [1ex]
        \hline
    \end{tabular}
    \medskip
    \begin{minipage}{\linewidth}\footnotesize
    Notes: 
    \par
    $^{a}$: the outer disk radius $R_{\rm out}$ is in units of the apparent inner disk radius $r_{\rm in}$; the physical inner disk radius $R_{\rm in} \approx 1.19 r_{\rm in}$ \citep{Kubota_1998}. 
    \par
    $^{b}$: disk normalization $N_{\rm disk}$ and inner disk radius $R_{\rm in}$ are related by $R_{\rm in} \sqrt{\cos\theta} \approx 1.19 \sqrt{N_{\rm disk}} d_{10}$, where $d_{10} = 13,900$ is the distance in units of 10 kpc. 
    \par
    $^{c}$: $N_{\rm apec}$ defined here as $10^{-5}$ times the {\sc xspec} {\fontfamily{pcr}\selectfont \emph {apec}} model normalization.
    \par
    $^{d}$: absorbed flux in the 0.3--10 keV band.
    \par
    $^{e}$: unabsorbed isotropic luminosity in the 0.3--10 keV band. 
    \end{minipage}
    \label{tab:diskir}
    }
\end{table}

\begin{figure*}
    \includegraphics[width=1.07\columnwidth]{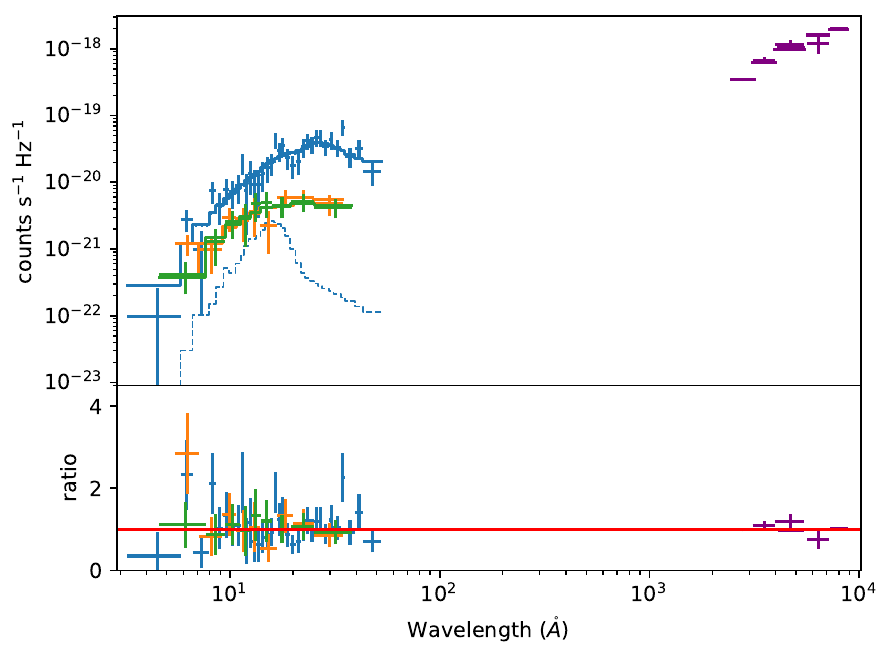}\hfill
    \includegraphics[width=1.07\columnwidth]{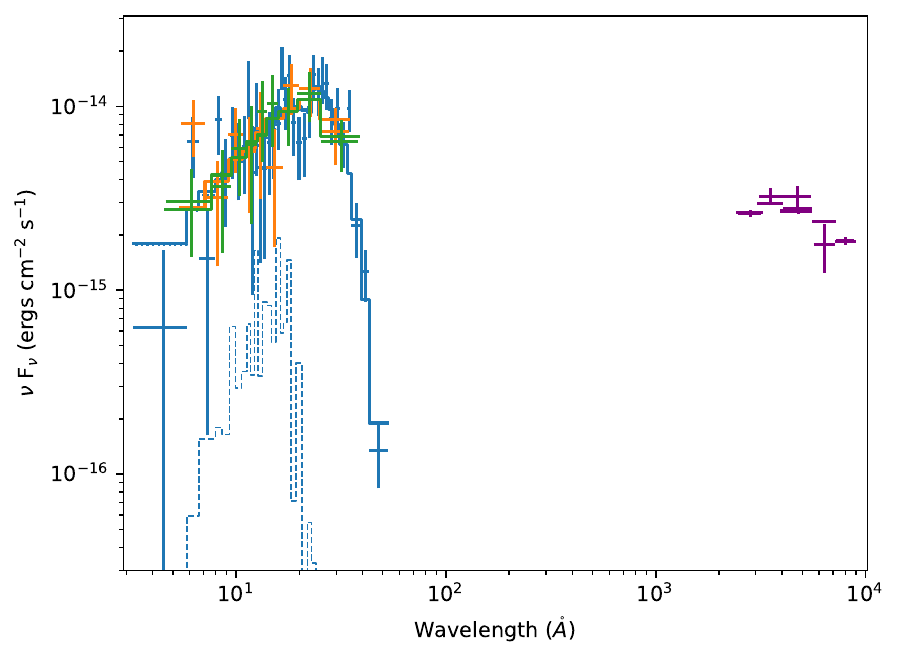}\\
    \includegraphics[width=1.07\columnwidth]{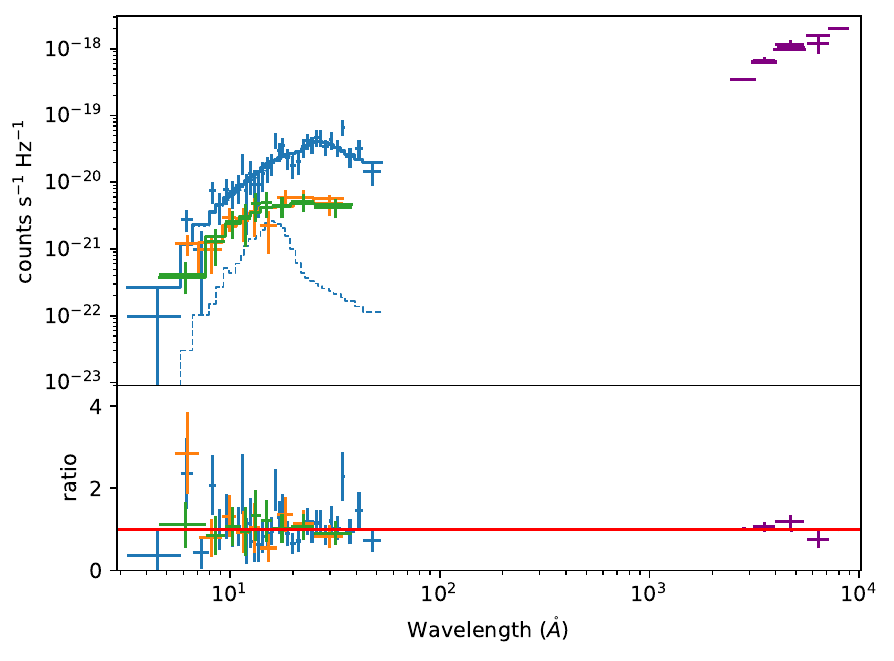}\hfill
    \includegraphics[width=1.07\columnwidth]{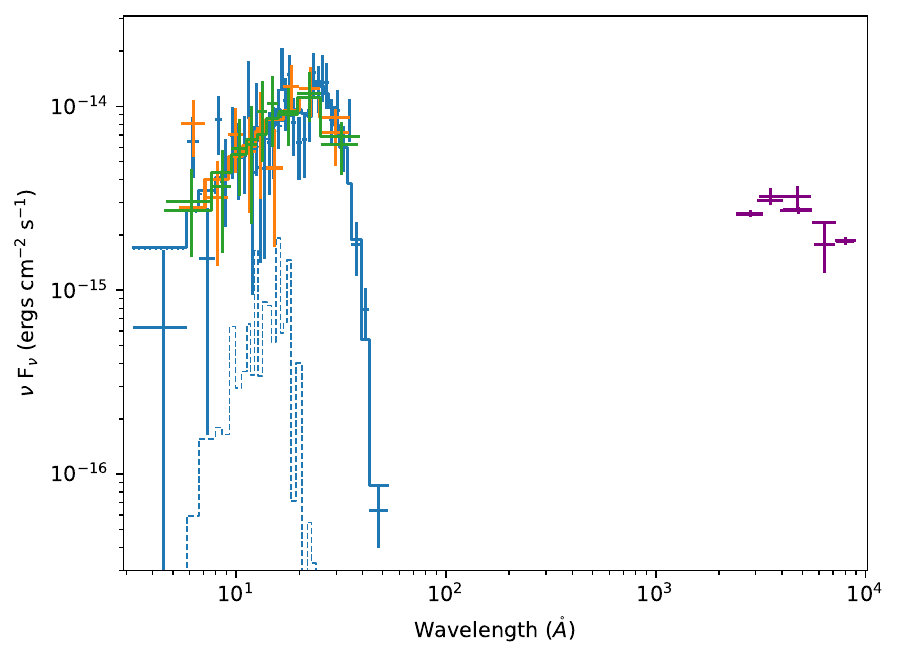}
    \caption{Combined X-ray (2023) and optical (2022--2023) data, fitted with alternative models. Colors are: blue for EPIC-pn; orange for EPIC-MOS1; green for EPIC-MOS2; purple for CFHT and {\it HST}. For the EPIC-pn data, we also plotted the modelled contribution from the contaminating X-ray emission of NGC\,6099 (dashed line). The left column contains the SEDs with the data/model ratios plotted underneath; the right column shows the unfolded spectra. The first row illustrates an irradiated accretion disk model (Section~\ref{section:diskir} and Table~\ref{tab:diskir}); the second row a combined model of a Comptonized X-ray disk plus optical/UV blackbody (Section~\ref{section:bbodyrad} and Table~\ref{tab:bbodyrad}).
    }
    \label{fig:XMM2_SED}
\end{figure*}

\subsection{Irradiated accretion disk}
\label{section:diskir}
To reduce the number of simultaneously free parameters, we fitted the SED with {\fontfamily{pcr}\selectfont \emph{diskir}} in three steps.
\par
First (as reported in Section~\ref{section:XMM2_obs}), we included only the X-ray datapoints, left the Comptonization fraction ($L_{\rm c}/L_{\rm d}$), photon index ($\Gamma$), peak disk temperature ($kT_{\rm in}$), disk normalization (proportional to $r_{\rm in}^2$) and intrinsic absorption column ($N_{\rm H,int}$) as free parameters, and froze the (unconstrained) electron temperature ($T_e$), irradiation radius ($r_{\rm irr}$), fraction of luminosity in the Compton tail thermalized in the inner disk ($f_{\rm in}$), reprocessing factor ($f_{\rm out}$) and outer disk size ($\log(r_{\rm out}/r_{\rm in})$) (the latter two parameters are unconstrained without optical/UV data). 
As in all other cases, we also included the small, fixed contamination component from the thermal plasma emission of NGC\,6099. 
\par
In the second step, we added the six optical/UV data points. 
We froze the X-ray parameters at their best-fitting value, thawed $f_{\rm out}$ and $\log(r_{\rm out}/r_{\rm in})$, and included two optical/UV extinction components ({\fontfamily{pcr}\selectfont \emph {redden} $\times$ \emph{redden}$_{\rm int}$}), one for the line-of-sight reddening (fixed at $E(B-V) = 0.05$ mag) and the other for the intrinsic reddening (keeping in mind that X-ray photons and optical photons come to us along different lines of sight and may see different amounts of scattering and absorption). 
Thus, in the second step, the fitting procedure has three free parameters available for six optical/UV data points.
\par
Finally, we thawed the X-ray parameters again and used the {\fontfamily{qcr}\selectfont steppar} command in {\sc xspec} around the previous best-fit values, to check whether the inclusion of the optical/UV part of the spectrum leads to small changes in the best-fitting values of the X-ray parameters. 
We found that such changes are negligible, that is the values of the X-ray parameters in the final SED model remain within the error range of the initial X-ray-only fit.
\par
The final model (Figure~\ref{fig:XMM2_SED}, Table~\ref{tab:diskir}) has $\chi^2 = 41/50$ dof. The best-fitting parameters related to the X-ray data have already been mentioned in Section~\ref{section:XMM2_obs}.  
The addition of the optical data enables us to determine three more parameters: the outer disk radius $R_{\rm out}$, the X-ray reprocessing fraction $f_{\rm out}$ and the intrinsic optical reddening $E(B-V)$. 
The outer disk radius $R_{\rm out} = (6 \pm 4) \times 10^{13} \,(\cos \theta)^{-1/2}$ cm.
The reprocessing fraction $f_{\rm out} \approx$ a few per cent is consistent with the range of values observed in ULXs \citep{grise12,Sutton_2014,Qiu_2021ApJ...922...91Q} and in the IMBH candidate ESO 243-49 HLX-1 \citep{soria17}. 
Finally, the best-fitting intrinsic optical reddening is low, consistent with line-of-sight only. 

\subsection{Stellar Cluster}
\label{section:bbodyrad}
Instead of a single irradiated disk model for the optical-to-X-ray data, we tried fitting the X-ray and optical data with separate, independent components: a Comptonization model ({\fontfamily{pcr}\selectfont \emph {simpl}}$\times${\fontfamily{pcr}\selectfont \emph {diskbb}} and a blackbody model ({\fontfamily{pcr}\selectfont \emph {bbodyrad}}), respectively. 
The best-fitting model (Figure~\ref{fig:XMM2_SED}, Table~\ref{tab:bbodyrad}) has $\chi^2 = 39/50$ dof, with an unabsorbed 0.3--10 keV luminosity $L_{\rm X} = (1.6 \pm 0.2) \times 10^{41}$ erg s$^{-1}$. 
The best-fitting parameters for the X-ray component are those already mentioned in Section~\ref{section:XMM2_obs}. 
The best-fitting temperature of the optical/UV blackbody component is $kT_{\rm bb} \approx 1.9$ eV  ({\it i.e.}, $T_{\rm bb} \approx 21,800$ K), consistent with a young stellar population dominated by OB stars. 
The best-fitting blackbody radius is $R_{\rm{bb}} = (2.8^{+0.5}_{-0.9}) \times 10^{13}$ cm.
\par
In summary, both the model in which the optical emission comes from an X-ray irradiated disk, and the model in which it comes from a stand-alone blackbody component lead to equally acceptable fits.
Not surprisingly, the same result is obtained also with a combination of disk and blackbody, for example a {\fontfamily{qcr}\selectfont \emph {diskir}} model with negligible X-ray reprocessing ($f \approx 0$) plus a blackbody.
The only significant difference between the two types of models is that an independent blackbody allows for a larger range of optical reddening, because temperature and normalization of the optical/UV component are not linked to the X-ray component.
For low reddening (total $E(B-V) \lesssim 0.1$ mag), the bolometric luminosity of the optical/UV component at the distance of NGC\,6099 is $\sim$10$^{40}$ erg s$^{-1}$. 
For higher reddening $E(B-V) \approx$0.3--0.6 mag (Table~\ref{tab:bbodyrad}), the intrinsic luminosity of the optical/UV emission can be as high as $\sim$10$^{41}$ erg s$^{-1}$.

\begin{table}
    \caption{
    Optical/X-ray spectral energy distribution from the 2022--2023 data, fitted with an X-ray Comptonization model plus an independent optical/UV blackbody: 
    \(\fontfamily{pcr}\selectfont \emph{TBabs} \times \emph{TBabs}_{int} \times (\emph{simpl} \times \emph{diskbb} ) + \emph {redden} \times \emph{redden}_{int} \times \emph{bbodyrad} \left[+ 0.04 \times  \emph{TBabs} \times  \emph{TBabs}_{\rm gal} \times \emph{apec}_{\rm gal}\,\right]\).
    The last term accounts for the contamination of thermal emission from the host galaxy NGC\,6099; we did not include this term in the flux and luminosity of HLX-1. Errors are 90\% confidence limits for one interesting parameter. The X-ray line-of-sight absorption is $N_{\rm H} \equiv 4.16\times 10^{20}$ cm$^{-2}$; the line-of-sight reddening is $E(B-V) \equiv 0.05$ mag.
    }
    \centering
    {
    \begin{tabular}{lccc} % three columns, alignment for each
        \hline
        Component & Parameter & Value & Unit\\
        \hline    
        \(\fontfamily{pcr}\selectfont \emph{redden}_{int}\) & $E(B-V)$  & $0.36_{-0.12}^{+0.18}$  & mag  \\ [1ex]
        \(\fontfamily{pcr}\selectfont \emph{TBabs}_{int}\) & $N_{\rm H, int}$ & $0.17_{-0.10}^{+0.05}$ & $10^{22}$ cm$^{-2}$ \\ [1ex]
        \multirow{3}{*}{\fontfamily{pcr}\selectfont \emph{simpl}} & $\Gamma$ &  $3.6_{-0.3}^{+0.9}$ &  -- \\ [1ex]
         & FracSctr & $0.18_{-0.07}^{+0.79}$ & -- \\ [1ex]
         & UpSc & [0.0] & -- \\ [1ex]
        \multirow{2}{*}{\fontfamily{pcr}\selectfont \emph{diskbb}} & $kT_{\rm in}$  & $0.07_{-0.04}^{+0.01}$   & keV \\ [1ex]
         & $N_{\rm disk}^{a}$ & $709_{-432}^{+1110}$ & km$^{2}$ \\ [1ex]
        \multirow{2}{*}{\fontfamily{pcr}\selectfont \emph{bbrad}} & $T_{bb}$ & $ 2.2_{-0.6}^{+3.2}$ & $10^{4}$ K\\ [1ex]
         & $N_{\rm bb}$ & $4.1_{-2.2}^{+1.4}$ & $10^{8}$ km$^{2}$\\ [1ex]
        \(\fontfamily{pcr}\selectfont \emph{TBabs}_{\rm gal}\) & $N_{\rm H,gal}$ & [0.57] & $10^{22}$ cm$^{-2}$  \\ [1ex]
        \multirow{2}{*}{$\fontfamily{pcr}\selectfont \emph {apec}_{\rm gal}$} & $kT$ & [0.37] & keV\\ [1ex]
         & $N_{\rm apec}^{b}$ & [6.85] & See Notes\\ [1ex]
        \hline
         & $\chi^2$/dof & 39/50 & \\ [1ex]
        \hline 
         & \(R_{\rm in} \sqrt{\cos\theta}\) & $44_{-16}^{+27}$  & $10^{9}$ cm\\ [1ex]
         & $R_{\rm bb}$  & $2.83_{-0.78}^{+0.51}$  & $10^{13}$ cm\\ [1ex]
         & $F_{\rm X}^{c}$ &  $1.4_{-0.1}^{+0.1} \times 10^{-14}$ & erg cm$^{-2}$ s$^{-1}$ \\ [1ex]
         & $L_{\rm X}^{d}$  & $1.6_{-0.2}^{+0.2} \times 10^{41}$ & erg s$^{-1}$\\ [1ex]
        \hline
    \end{tabular}
    \medskip
    \begin{minipage}{\linewidth}\footnotesize
    Notes: 
    \par
    $^{a}$: disk normalization $N_{\rm disk}$ and inner disk radius $R_{\rm in}$ are related by $R_{\rm in} \sqrt{\cos\theta} \approx 1.19 \sqrt{N_{\rm disk}} d_{10}$, where $d_{10} \approx 13900$ is the distance in units of 10 kpc. 
    \par
    $^{b}$: $N_{\rm apec}$ defined here as $10^{-5}$ times the {\sc xspec} {\fontfamily{pcr}\selectfont \emph {apec}} model normalization.
    \par
    $^{c}$: absorbed flux in the 0.3--10 keV band.
    \par
    $^{d}$: unabsorbed isotropic luminosity in the 0.3--10 keV band. 
    \end{minipage} 
    \label{tab:bbodyrad}
    }
\end{table}

\section{Discussion}
The observed properties of HLX-1 make it an unusual and intriguing source, difficult to categorize into ordinary classes of accreting compact objects. 
All three X-ray detections ({\it Chandra} in 2009, {\it XMM-Newton} in 2012 and 2023) show a soft, likely thermal spectrum. 
However, the spectra are not consistent with the ``canonical'' high/soft state with constant normalization (standard disk truncated at the innermost stable circular orbit) and $L_{\rm disk} \propto T_{\rm in}^4$. 
The uncertainty in the interpretation of the inner disk radius across different epochs and for different models, together with the lack of optical redshift measurements and the lack of simultaneous optical/X-ray observations, make it more difficult to distinguish between IMBH scenarios at the distance of NGC\,6099 and background AGN.
Nonetheless, we can assess pros and cons of several alternative scenarios. 
\par
\subsection{Canonical state transitions of an IMBH?}
\label{section:Canonical_state_transitions}
In this scenario, HLX-1 was in a canonical sub-Eddington state in 2009, a super-Eddington state in 2012, and back to a sub-Eddington state in 2023.
This requires an IMBH accretor with a mass $\sim$10$^{3}$--10$^{4} M_\odot$.
We showed (Section~\ref{section:combinefit}) that the 2009 and 2023 spectra are consistent with the same state, and suggest a characteristic radius $R_{\rm in} \sqrt{\cos\theta} \approx 10,000$ km. 
If $R_{\rm in}$ corresponds to the innermost stable circular orbit $R_{\rm isco}$, the BH mass is a few $\times$ 10$^{3} M_\odot$ (with the exact value depending on the unknown spin parameter), self-consistent with the luminosity argument ($L_{\rm Edd} \sim$ a few $\times 10^{41}$ erg s$^{-1}$).
\par
Standard disk instability models predict that below a threshold accretion rate (or, equivalently, for an outer disk larger than a minimum radius), the outer accretion disk becomes mostly neutral and may give rise to thermal-viscous instability cycles \citep[{\it e.g.},][]{king97,burderi98,dubus99,dubus01,lasota01,kalogera04}. 
Such cycles appear observationally as X-ray state transitions. 
To explore this scenario, let us assume that the optical emission of HLX-1 comes mostly from a large accretion disk. 
At the large outer radius inferred from the {\fontfamily{qcr}\selectfont \emph {diskir}} model fit, $R_{\rm out} = (6 \pm 4) \times 10^{13} \,(\cos \theta)^{-1/2}$ cm (Section~\ref{section:diskir}), the source of optical emission is mostly X-ray irradiation and reprocessing, rather than viscous dissipation. 
Based on the irradiation temperature profiles calculated by \cite{dubus99}, the critical accretion rate below which the transient behaviour occurs is 
\begin{equation}
   \dot{M}_{\rm cr} \approx 3.2 \times 10^{21} \left( \frac{M}{10^4 M_\odot}  \right)^{-0.4} \left( \frac{R_{\rm out}}{6 \times 10^{13}\, {\mathrm {cm}}}  \right)^{2.1} \ {\mathrm {g~s}}^{-1}
\end{equation}
The accretion rate corresponding to the luminosity inferred in 2009 and 2023 is $\dot{M} \approx L_{\rm X}/(0.1 c^2) \approx 10^{21}$ g s$^{-1}$. 
Thus, it is possible that such accretion rate is not sufficient to keep hydrogen fully ionized in the outer disk, giving rise instead to transient cycles. 
It is difficult to assess the plausibility of this scenario, because it is not common behaviour for stellar-mass accretors (let alone putative IMBH accretors) to switch between a canonical high/soft state and a highly super-Eddington state without transiting through other states.
Future X-ray observations of HLX-1 for example in the low/hard state are needed to constrain and support or refute this scenario. 
However, we may not have a chance to see such a transition to the low/hard state happen in our lifetime, because the characteristic outburst decay timescale ({\it{i.e.},} the time required for the outside-in propagation of a cooling front) for a disk radius of $6 \times 10^{13}$ cm and a BH mass of $10^4 M_\odot$ is $\approx$300 yr \citep{hameury20}. 
\par
An additional issue to consider is whether a single donor star can steadily feed the IMBH accretion disk at the required rate, via Roche lobe overflow, without being tidally disrupted. 
Assuming a mass ratio $q \sim 10^{-3}$ between donor star and IMBH, an accretion disk radius $R_{\rm out} \approx 6 \times 10^{13}$ cm suggests a binary separation $a \approx 10^{14}$ cm and a Roche lobe radius $R_2 \approx 70 R_\odot$ for the donor star \citep{paczynski77,eggleton83,frank02}. 
Thus, in principle, there are several types of young (blue/yellow supergiant) or old (red giant) stars that can fill their Roche lobe in a circular orbit around an IMBH. 
The expected mass transfer rate depends on how the internal structure of the donor star responds to rapid mass loss, and how its orbit widens or shrinks under the combined effect of angular momentum redistribution and gravitational decay between the star and the IMBH \citep{webbink85,hjellming87,ge10,dai13a,dai13b}. 
Further investigation of this issue is beyond the scope of this paper.

\par
An alternative scenario we shall only briefly mention here is that the IMBH may instead be recurrently fed at periastron by a donor star on an eccentric orbit, possibly even undergoing partial tidal disruption at each periastron passage \citep{chen21,nixon21,Cufari_2022}.  
In this scenario, the 2012 observation was taken shortly after a periastron flare, while the 2009 and 2023 observations were taken later in a flare decline phase. 
A similar scenario was proposed and investigated to explain the X-ray outbursts in the other strong IMBH candidate ESO\,243-49 HLX-1 \citep{godet14,vanderhelm16} and in some galactic nuclear transients \citep{campana15,Liu_2023}. 

\par
Finally, regardless of the physical reason for the outbursts, if the accreting IMBH exceeds its Eddington luminosity, we expect the fitted radius of the thermal X-ray emission to become larger than the innermost stable circular orbit. 
This is because in super-critical accretion, such large radii correspond to the spherization radius \citep{Shakura&Sunyaev_1973,poutanen07}, $R_{\rm sph} \propto \dot{m} R_{\rm in}$, the radius at which the disk becomes geometrically thick, advective, and from where fast outflows are launched. 
The situation may be more complicated, if the thermal continuum is not the optically thick emission from a section of the disk surface, but instead emission from the scattering photosphere of the wind, for example the walls of a polar funnel; this is still an unsolved question even in nearby ULXs with much better observational coverage \citep[{\it e.g.},][]{narayan17,walton20,robba21,gurpide21,barra22,barra24,walton24}.

\subsection{A full TDE with delayed X-ray peak?}
In this scenario, the similar X-ray flux level and soft spectrum seen in 2009 and 2023 (before and after the 2012 peak) is only a coincidence and does not correspond to a similar intrinsic luminosity and spectral state in the two epochs.
The 2009 spectrum corresponds to the early rise of the TDE emission, the 2012 spectrum was taken near the peak, and the 2023 data come from the decline phase.
\par
After two decades of intense theoretical and observational studies, it is still actively debated what fraction of a TDE emission originates from viscous dissipation and accretion onto the BH, and what fraction from shocks (stream-stream collisions) during the debris circularization phase \citep[{\it{e.g.}},][]{lodato11,piran15,metzger16,metzger22,steinberg24}. 
\par
The first possibility is that in 2009, HLX-1 was in the initial super-Eddington accretion phase, and the accretion rate has monotonically declined since then. 
The main feature of a super-Eddington phase is the launching of thick outflows, which downscatter and reprocess the X-ray photons emitted close to the compact object \citep[{\it{e.g.}},][]{lodato11,roth16,metzger16,dai18,metzger22,thomsen22,bu22}. 
The initial spectrum peaks in the optical/UV; soft X-rays rise later, after the super-Eddington outflow has stopped and the reprocessing envelope has cooled or shrunk \citep{metzger16,roth16,Chen_2018,Wevers_2019,metzger22}.
In particular, for an initially super-Eddington IMBH TDE, the time-scale $t_{\rm Edd}$ on which the fallback rate declines below the Eddington rate is $t_{\rm Edd} \simeq 14$ $\eta_{0.1}^{3/5}M_{4}^{-2/5} r_{\ast}^{3/5} m_{\ast}^{1/5}$ yr \citep{Chen_2018}, where $\eta_{0.1} \equiv \eta/0.1$ is the radiative efficiency, $M_{4}$ is the BH mass in units of $10^4 M_{\odot}$, $r_{\ast}$ and $m_{\ast}$ are the disrupted star's radius and mass in solar units.
Therefore, the initial optically bright, X-ray faint phase may last for several years in IMBH TDEs~\citep{Chen_2018,tang24}. 
If this is the correct timeline of events for HLX-1, it implies that in 2009, at least 99\% of the emitted X-ray photons were blocked by the reprocessing envelope, and, as a result, the optical/UV luminosity should have been at least as high as the X-ray luminosity in 2012 (a few times $10^{42}$ erg s$^{-1}$), corresponding to a visual magnitude $m_{g,{\rm AB}} < 19$ mag. 
Such optical brightening should have been detectable by the main wide-field transient surveys active circa 2009. 
We inspected 34 $R$-band images (exposure time of 60 s each) taken by the Palomar Transient Factory (PTF) \citep{rau09,law09} from the 48-in Samuel Oschin Telescope between 2009 May and 2010 August\footnote{Available from {\url{https://irsa.ipac.caltech.edu/applications/ptf}.}}. 
We did not find any source at the location of HLX-1 in any frame, down to an individual-frame detection limit $m_{R,{\rm AB}} \approx 20.5$ mag. 
We used the {\sc{iraf}} task {\fontfamily{qcr}\selectfont imcombine} to build separate stacked images of all the 2009 and 2010 observations, and obtained that HLX-1 is not detected in either year, down to a limit $m_{R,{\rm AB}} \approx 21.5$ mag.  
Further PTF observations from the 48-in telescope were also taken in 2013 February (5 60-s exposures), March (41 exposures), April (47 exposures) and May (46 exposures). 
We also built separate stacked images for those four epochs and verified once again a non-detection, with an upper limit $m_{R,{\rm AB}} \approx 21$ mag in 2013 February and $m_{R,{\rm AB}} \approx 21.5$ mag for the other three months. 
This suggests that shortly after the 2012 X-ray outburst, there was no bright optical/UV envelope, either. 
We conclude that the super-Eddington optical/UV reprocessing scenario is not supported by the optical data. 
\par
The second possibility is that the bolometric luminosity of HLX-1 was sub-Eddington and much lower than in 2012 (consistent with the lack of a bright optical counterpart). 
In this case, the X-ray emission in 2009 was due to shocks in the self-colliding stream or from stream compression near the pericentre, during the initial circularization phase of the TDE, before disk formation \citep{piran2015circularization,Wevers_2019,chen21,Liu_2022,huang24,steinberg24}.
Instead, the emission seen in 2012 and 2023 was from viscous dissipation and accretion, after debris circularization. 
This scenario is more plausible if there is a substantial time delay (months or years) between stellar disruption and disk formation. 
One of the parameters that affect the timescale for debris circularization is the amount of apsidal precession of the stream, which decreases at lower BH masses \citep{shiokawa15,guillochon15,dai15,hayasaki16,bonnerot17,Liu_2022}. 
For example, the candidate TDE ASASSN-15oi had a slow-rising phase in the soft X-ray band for about one year \citep{gezari17,holoien18} which may correspond to a slow circularization phase around a relatively light BH. 
The recently discovered soft X-ray transient EP240222a (Chichuan Jin et al., submitted)\ may have had a three-year circularization phase, consistent with an IMBH with $M \sim 10^5 M_\odot$. 
In summary, HLX-1 may be a good representative of the class of IMBH TDEs in globular clusters, in which faint precursor X-ray emission from stream-stream collisions occurs months or years before the peak of the accretion-powered emission. 

\subsection{A background changing-look, supersoft AGN?}
As anticipated in the outline of our source selection (Section~\ref{section:selection_criteria}), the two main X-ray properties (flux variability by two orders of magnitude and extreme softness), by themselves, are not fireproof evidence of a high-state IMBH.
We have already mentioned that some changing-look AGN \citep{Komossa_2023} show dramatic X-ray luminosity variations over a few years. 
A good example is 1ES 1927$+$654 \citep{gallo13,Ricci_2020,li22,Masterson_2022,cao23,li24} ($d \approx 87$ Mpc, $z \approx 0.019$), which is believed to have reached its Eddington limit at $L_{\rm X} \approx 10^{44}$ erg s$^{-1}$, with a soft, thermal spectrum and peak temperature $kT_{\rm in} \approx 0.15-0.20$ keV, during its well-monitored 2019--2020 outburst. 
The spiral galaxy IC\,3599 ($d \approx 93$ Mpc, $z \approx 0.021$) is another good example of changing-look AGN \citep{campana15,Grupe_2015,grupe24}: it showed at least 2 outbursts (1990 and 2010), reaching peak luminosities of a few $\times 10^{43}$ erg s$^{-1}$, from a baseline luminosity of a few $\times 10^{40}$ erg s$^{-1}$. 
IC\,3599 also shows soft spectra at all epochs, well fitted by a blackbody with $kT_{\rm bb} \approx 0.1$ keV or a steep power-law spectrum with $\Gamma > 3$ \citep{campana15}. 
A sample of 60 nuclear sources with a supersoft spectrum (power-law photon index $\Gamma \gtrsim 3$ and/or a dominant blackbody component at $kT_{\rm bb} \sim 0.1-0.2$ keV) was collected and discussed by \cite{sacchi23} from the 4XMM-DR9 catalogue. 
\par
One possible explanation for changing-look, soft-spectrum AGN is a TDE or partial TDE \citep{campana15,Ricci_2020,nixon21,chen21,Cufari_2022} on a previously active AGN (that is, with a pre-existing accretion disk). 
In other cases, such as the supersoft (pure thermal spectrum with $kT_{\rm bb} \approx 0.2$ keV) Seyfert 2 nucleus 2XMM J123103.2$+$110648 \citep{Terashima_2012,Lin_2013}, persistent but highly variable activity over many years suggests that it is not a TDE. 
Explaining the physics of changing-look and supersoft AGN is beyond the scope of this work. 
What matters here is whether we might have misidentified one such (background) AGN for an off-nuclear source in NGC\,6099. To answer this question, we need to look at its optical appearance. 
\par
In the 2022--2023 datasets, HLX-1 has a characteristic observed 0.3--10 keV flux over $u$-band flux $\approx$10 (Table ~\ref{tab:CFHT_info},~\ref{tab:params_list}); for the $r$ band, the X-ray over optical flux ratio is $\approx$ 20. 
Such values are low enough to be consistent with AGN and IMBHs. 
By comparison, the extreme changing-look AGN 1ES 1927$+$654 has $f_{\rm XO} \approx 10$ from pre-outburst {\it XMM-Newton} observations at $L_{\rm X} \approx 10^{43}$ erg s$^{-1}$ \citep{gallo13} and Pan-STARRS $i$-band images. 
The main constraint comes instead from the point-like, faint appearance of the optical counterpart. 
\par
The optical counterpart of 1ES 1927$+$654 (including only the host galaxy contribution) would look as faint as the optical counterpart of HLX-1 only if that galaxy was located at a luminosity distance 40 times higher than its real distance \citep{li22,li24}, that is $\approx$3.5 Gpc ($z \approx 0.58$)\footnote{For all distance and redshift calculations in this paragraph and the following two, we used Ned Wright's Cosmology Calculator \citep{wright06}.}.
If HLX-1 were a changing-look AGN at that distance, it would have an X-ray luminosity $L_{\rm X} \approx 2 \times 10^{45}$ erg s$^{-1}$ in 2012, and $L_{\rm X} \approx 2 \times 10^{43}$ erg s$^{-1}$ in 2009 and 2023, which is still within the range of plausible Seyfert luminosities.
We want to ascertain whether at such distance, the host galaxy would look point-like or extended. 
For this test, we take the $V$-band surface brightness profile for 1ES 1927$+$654 measured by \cite{li22} (their Fig.~1); we neglect K corrections, and assume for simplicity that the surface brightness scales as $(1+z)^{-4}$. 
We then estimate the R$_{25}$ radius at which the observed $V$-band surface brightness $\mu_V = 25$ mag arcsec$^{-2}$ for a galaxy identical to 1ES 1927$+$654 but located at $z \approx 0.58$. We obtain R$_{25}$ $\approx 0\farcs5$. 
Therefore, the stellar disk emission of such galaxy would be resolvable in {\it HST} images even at that redshift. 

\par
Let us consider instead the possibility that HLX-1 is a background AGN hosted by a dwarf galaxy, for example like the dwarf Seyfert 1 galaxy POX 52
\citep{Barth_2004, Thornton_2008}, located at a luminosity distance of 98 Mpc ($z = 0.0218$). 
POX 52 would need to be located 30 times further away, at $z \approx 0.50$ (luminosity distance $\approx$2.9 Gpc) to look as faint as our observed optical source. 
At that distance, we would measure an R$_{25}$ $\approx 0\farcs3$; thus, even in this case, it would look extended in {\it HST}. 
\par
Finally, the observed X-ray flux itself constrains the possible distance range of HLX-1 in the background AGN scenario. 
A distance higher than $z \approx 1.3$ is very unlikely, because it would push its rest-frame peak X-ray luminosity above $10^{46}$ erg s$^{-1}$, an approximate upper limit for the X-ray luminosity of AGN \citep{Singal_2022}. 
At that redshift, the upper limit of $0\farcs15$ to the optical size of HLX-1 (Section~\ref{section:photometry}) corresponds to $\approx$1.3 kpc. 
\par
In summary, the point-like appearance and optical faintness of HLX-1 are more consistent with the IMBH scenario in a globular cluster or ultracompact dwarf (UCD) in the NGC\,6098/6099 group rather than a background Seyfert galaxy. 

\subsection{Host star cluster or irradiated disk?}
Let us assume now that our source is indeed near NGC\,6099.
Then, its optical counterpart is either coming entirely from a young star cluster, or it is a mix of the (bluer) contribution from an accretion disk and a (redder) contribution from the stellar population in the star cluster. 
\par
If all the optical emission comes from the star cluster, the observed blue colors require a young stellar population.
We used the population synthesis code {\sc starburst99} \citep{Leitherer_1999, Leitherer_2014}, with the Geneva evolutionary tracks \citep{ekstrom12}, to quantify the range of acceptable ages and masses, at different metallicities.
From the optical brightness and colors measured from the CFHT and {\it HST} data (Section~\ref{section:photometry}), corrected for line-of-sight Galactic reddening, and assuming an impulsive star formation, we obtain an age of $\approx$6--8 Myr and stellar mass $M_{\ast} \approx 4 \times 10^4 M_{\odot}$ at solar metallicity ($Z = 0.014$), or an age of $\approx$20--30 Myr and stellar mass $M_{\ast} \approx 2.4 \times 10^5 M_{\odot}$ at very low metallicity ($Z = 0.002$).
Ages younger than $\approx$6 Myr are ruled out by the moderately red $V-I$ colour ($V-I \approx (0.5 \pm 0.2)$ mag, Section~\ref{section:photometry}), which points to the minimum age at which the most massive surviving stars of the cluster evolve to the red supergiant stage instead of dying as blue stars.
Instead, if we include an intrinsic reddening $E(B-V) \approx$0.3--0.6 mag (Table~\ref{tab:bbodyrad}), we can explain the optical/UV emission with a young stellar population of mass $\approx$10$^5 M_\odot$ and ages $\lesssim$5 Myr.
\par
The young cluster scenario faces several challenges.
IMBHs are indeed predicted to form in the core of young clusters via core collapse and stellar collisions, on timescales of few Myr \citep{Portegies_Zwart_2002, COLEMAN_MILLER_2004, Freitag_2006}.
However, numerical simulations by \cite{Di_Carlo_2021} with the {\sc mobse} population synthesis code \citep{Mapelli_2017} suggest that clusters with stellar masses up to $3 \times 10^4 M_{\odot}$ (similar to our solar-metallicity case) do not form IMBHs more massive than $\approx$500 $M_{\odot}$, insufficient to explain the observed luminosity of HLX-1.
Moreover, there is no sign of recent star formation (for example other young star clusters) in the halo of NGC\,6099.
We cannot exclude that the young star cluster with its central IMBH came from a gas-rich satellite dwarf, accreted and disrupted by NGC\,6099; however, we do not see tidal tails or other signatures of such a recent event.
\par
The second possibility is that the bluer component of the optical emission comes from the X-ray-irradiated disk or, more generally, the irradiated accretion inflow/outflow, and the host stellar cluster contains only an old, red population (much fainter for the same total stellar mass).
For example, with {\sc starburst99} we find that an old globular cluster or UCD with an age of $\approx10$ Gyr is consistent with the observed $I$-band luminosity (and obviously also with the brightness measured in bluer bands) for a stellar mass up to $\approx$2 $\times 10^7 M_{\odot}$.
From the relation between central BH mass and stellar mass valid for UCDs and stripped nuclear star clusters \citep{Graham_2023,mayes24}, a stellar system of this mass can harbour a BH of up to $\sim$10$^6 M_{\odot}$.
Even a cluster with a much lower stellar mass $\approx$10$^6 M_{\odot}$ may harbor an IMBH up to $\sim$10$^4 M_{\odot}$ (as may be the case for the Galactic globular cluster Omega Cen: \citealt{haberle24}), sufficient to explain the X-ray luminosity of our source.
An old star cluster also implies that the only plausible feeding mechanism is tidal stripping or disruption of a low-mass star.
\par
The apparent optical radius $R_{\rm out}=(6\pm4)\times10^{13}$ cm (from our {\fontfamily{qcr}\selectfont \emph {diskir}} model fitting; Section~\ref{section:diskir}) is an order of magnitude larger than the predicted circularization radius $R_{\rm c}$ of a low-mass main-sequence star ($R_{\rm c} \approx 2 R_{\rm t} \approx 2 R_{\ast}\, \left( M_{\rm BH}/M_{\odot} \right)^{1/3} \sim 40 R_{\odot} \approx 3 \times 10^{12}$ cm, where $R_{\rm t}$ is the tidal disruption radius, $R_{\ast}$ is the radius of main-sequence star, and where we assumed $M_{\rm BH} = 10^4 M_{\odot}$). 
Large optical radii are a well-known problem in TDE models ({\it e.g.}, \citealt{Gezari_2021}, particularly evident in her Fig.~8), where there is a significant discrepancy of one and sometimes two orders of magnitude between the fitted blackbody radii and the expected circularization radii of the debris disks. 
This could be due to viscous spreading of the gas in the disk, inwards and outwards \citep{frank02,vanVelzen_2019ApJ...878...82V}.
Alternatively, the self-intersection radius of the debris stream should be taken as the most realistic scale for the disk size \citep{Gezari_2021}. For $M_{\rm BH} \sim 10^4 M_{\odot}$, such radius is $\approx$ a few $\times 10^{13}$ cm \citep{dai15}, consistent with the apparent optical radius in HLX-1.
\par
The relative fraction of optical emission from the irradiated disk (function of $L_{\rm X}$) and the host star cluster (constant) can only be determined from repeated optical observations at different X-ray luminosities, and deeper observations in the near-IR.
For example, the observed long-term decrease in the blue-band luminosity after X-ray outbursts in the IMBH candidate ESO 243-49 HLX-1 proved that at least the bluer component of the optical emission was coming from X-ray-irradiated gas in the inflow/outflow \citep{Farrell_2014,soria17}.

\section{Conclusions}
We identified an intriguing high-luminosity, point-like X-ray source from {\it{XMM-Newton}} and {\it{Chandra}} archives. The source satisfies the main selection criteria for an IMBH.
It appears located at the outskirts of the elliptical galaxy NGC\,6099 ($d \approx 139$ Mpc); we called it NGC\,6099 HLX-1.
The source was detected by {\it Chandra} and {\it XMM-Newton} at different flux levels in three separate epochs (lower fluxes in 2009 and 2023, highest flux in 2012).
Its peak luminosity ($L_{\rm X} \approx$ a few $\times10^{42}$ erg s$^{-1}$) combined with a consistently soft X-ray spectrum (optically thick thermal component with $kT_{\rm in} \approx 0.2$ keV plus power-law component with photon index $\Gamma \gtrsim 3$) are predicted hallmarks of IMBHs near or above their Eddington limit.
Luminosity, time evolution, and spectral properties rule out highly beamed or highly super-Eddington stellar-mass accretors, or a young supernova.
On the other hand, X-ray variability and soft thermal spectra are also seen in some TDEs and changing-look AGN.
\par
Moreover, we discovered a blue, point-like optical counterpart in CFHT images; it is also unresolved in follow-up {\it HST} images.
We estimate a brightness $V_0 = (24.6 \pm 0.2)$ mag (corrected for line-of-sight Galactic reddening), an absolute magnitude $M_{V,{\rm Vega}} = (-11.1 \pm 0.2)$ mag and optical colors (corrected for line-of-sight Galactic reddening) $B-V \approx -0.1$ mag, $V - I \approx 0.5$ mag (Vegamag system).
The morphology and brightness of the optical counterpart are consistent with a source at the outskirts of NGC\,6099, such as a massive star cluster or UCD. 
We cannot completely rule out a distant background AGN or quasar. 
However, the latter explanation is more contrived, given the lack of any spatially extended optical feature around the X-ray source. 
If the AGN was so far away that its host galaxy is undetectable, its rest-frame X-ray luminosity would be unphysically high.
There is no evidence of tidal tails or recent star formation in the halo of NGC\,6099: thus, we suggest that the IMBH is more likely surrounded by an old or intermediate-age stellar population, and that the blue optical emission comes mostly from the X-ray-irradiated accretion flow.
A definitive answer for the nature of the optical counterpart will only come from follow-up optical spectroscopy (for the redshift) and photometry (to search for evolution of the optical colors as a function of X-ray flux).  
\par
We discussed alternative interpretations, and argued that an IMBH in a compact star cluster, fed by tidal stripping or tidal disruption of a low-mass star, is the simplest explanation consistent with the data at hand.
If so, the obvious question is then why the source was already seen in a moderately bright, soft X-ray state in 2009, three years before the 2012 highest luminosity state.
At first sight, the 2009 detection seems to rule out a single TDE.
One possible answer is that HLX-1 is fed by tidal stripping of a companion star on an eccentric orbit (partial tidal disruption).
This is a model suggested for example to explain the repeated X-ray outbursts in the best-known IMBH candidate ESO 243-49 HLX-1.
An alternative scenario is that the 2009 observation corresponds to the initial rising phase of a TDE, when the thermal X-ray emission comes mostly from shocked gas in the self-intersecting accretion stream; instead, the 2012 observation ($\sim$50--100 times more luminous, depending on the choice of spectral models) corresponds to the disk accretion phase.
Follow-up X-ray observations will be needed to determine whether the X-ray source is now steadily declining along the expected TDE track (luminosity $\propto t^{-5/3}$), whether and at what luminosity threshold it will switch to the low/hard state (which will constrain the BH mass), or, conversely, whether it will rise again in the future, if the feeding source was not completely disrupted.

\section*{Acknowledgements}
We thank the anonymous referee for their detailed corrections and suggestions, which led to a much improved paper. 
We thank Roberta Amato, Lixin Dai, Hua Feng, Miroslav Filipovic, Andres Gurpide, Chichuan Jin, Mansi Kasliwal, Ji-Feng Liu, Matt Middleton, Rong-Feng Shen, Beverly Smith for fruitful discussions. YCC acknowledges hospitality and support from the Osservatorio Astrofisico di Torino during part of this work. RS acknowledges hospitality and support from the National Astronomical Observatories of China (University of the Chinese Academy of Sciences, Beijing). This work was performed in part at the Aspen Center for Physics, which is supported by National Science Foundation grant PHY-2210452.
This project is supported by the National Science and Technology Council of the Republic of China (Taiwan) through grants 111-2112-M-007-020 and 112-2112-M-007-042. 
RS acknowledges the INAF grant number 1.05.23.04.04 and also acknowledges the grant number 12073029 from the National Science Foundation of China.
IC's research is supported by the Telescope Data Center, Smithsonian Astrophysical Observatory. 
This work is partly based on observations obtained with {\it XMM-Newton}, an ESA science mission with instruments and contributions directly funded by ESA Member States and NASA.
We used data obtained from the {\it Chandra} Data Archive and the {\it Chandra} Source Catalog, contained in~\dataset[doi: 10.25574/cdc.279]{https://doi.org/10.25574/cdc.279}, and software provided by the {\it Chandra} X-ray Center (CXC) in the {\sc ciao} application package. 
This research is also partly based on observations made with the NASA/ESA {\it Hubble Space Telescope}, obtained from the Mikulski Archive for Space Telescopes (MAST) at the Space Telescope Science Institute, which is operated by the Association of Universities for Research in Astronomy, Inc., under NASA contract NAS 5–26555. The specific observations analyzed can be accessed via~\dataset[doi: 10.17909/qfmj-4j20]{http://dx.doi.org/10.17909/qfmj-4j20}. These observations are associated with program ID HST-SNAP-17177. Support for program HST-SNAP-17177 was provided by NASA through a grant from the Space Telescope Science Institute
Furthermore, we used data from the European Space Agency (ESA) mission
{\it Gaia} (\url{https://www.cosmos.esa.int/gaia}), processed by the {\it Gaia} Data Processing and Analysis Consortium (DPAC,
\url{https://www.cosmos.esa.int/web/gaia/dpac/consortium}). Funding for the DPAC has been provided by national institutions, in particular the institutions participating in the {\it Gaia} Multilateral Agreement.
We used {\sc iraf} software for part of the optical analysis: {\sc iraf} is distributed by the National Optical Astronomy Observatory, which is operated by the Association of Universities for Research in Astronomy (AURA) under a cooperative agreement with the National Science Foundation. We also used the Vizier data archive and the Aladin sky atlas developed and maintained by CDS, Strasbourg Observatory, France.

%baumann22 = https://ui.adsabs.harvard.edu/abs/2022ASPC..532....7B/abstract
%\end{acknowledgments}


\vspace{5mm}
\facilities{{\it{XMM-Newton}}, {\it{Chandra X-ray Observatory}}, {\it{Neil Gehrels Swift Observatory}}, {\it{Hubble Space Telescope}}, Canada France Hawaii Telescope}


\software{
 \texttt{AstroImageJ} \citep{Collins_2017},
 \textsc{CIAO} \citep{Fruscione_2006},
 \texttt{Sas} \citep{2004ASPC..314..759G},
 \texttt{Sherpa} \citep{Freeman_2001, 2024ApJS..274...43S, doug_burke_2022_7186379},
 \textsc{IRAF} \citep{1986SPIE..627..733T,1993ASPC...52..173T},
 \texttt{ftools} \citep{Blackburn_1995, NASA_2014},
 \textsc{xspec} \citep{Arnaud_1996}
 }

\onecolumngrid

\bibliography{ngc6099_hlx-1}{}
\bibliographystyle{aasjournal}

\end{document}